\documentclass[fleqn,usenatbib]{mnras}
\usepackage[T1]{fontenc}
\DeclareRobustCommand{\VAN}[3]{#2}
\let\VANthebibliography\thebibliography
\def\thebibliography{\DeclareRobustCommand{\VAN}[3]{##3}\VANthebibliography}

\newcommand{\aznum}{123~} 

\usepackage{graphicx}	
\usepackage{amsmath}	
\usepackage{threeparttable}
\usepackage{multirow}
\usepackage{threeparttablex}
\usepackage{longtable}
\usepackage{makecell}
\usepackage{bm}
\usepackage{amssymb}

\title[Constraints from Fermi observations of LGRBs]
{Constraints from Fermi observations of Long Gamma-Ray Bursts on cosmological parameters}

\author[Wang and Liang]{
Huifeng Wang,$^{1}$
Nan Liang,$^{1,2,3}$\thanks{E-mail:liangn@bnu.edu.cn}
\\
$^{1}$Key Laboratory of Information and Computing Science Guizhou Province, Guizhou Normal University, Guiyang, Guizhou 550025, China\\
$^{2}$School of Cyber Science and Technology, Guizhou Normal University, Guiyang, Guizhou 550025, China\\
$^{3}$Joint Center for FAST Sciences Guizhou Normal University Node, Guiyang, Guizhou 550025, China\\
}

\date{Accepted XXX. Received YYY; in original form ZZZ}

\pubyear{2024}

\begin{document}

\label{firstpage}
\pagerange{\pageref{firstpage}--\pageref{lastpage}}
\maketitle
\begin{abstract}
In this paper, we compile a \emph{Fermi} sample of the \emph{long} GRB 
observations from 15 years of GBM catalogue with identified redshift, in which the GOLD sample contains 123 long GRBs at $z\le5.6$ and the FULL sample contains 151 long GRBs with redshifts at $z\le8.2$.
The Amati relation (the $E_{\rm p,i}$-$E_{\rm iso}$ correlation) are calibrated at $z<1.4$ by a Gaussian Process from the latest observational Hubble data (OHD)  with the cosmic chronometers method  so that GRBs at high-redshift $z\ge1.4$ can be used to constrain cosmological models via the Markov chain Monte Carlo (MCMC) method.
From the cosmology-independent GRBs with the GOLD sample at $z\ge1.4$ and the Pantheon+ sample of type Ia supernovae (SNe Ia) at $0.01<z\leq2.3$, we obtain $\Omega_{\rm m} = 0.354\pm0.018, H_0 = 73.05\pm0.2\,\rm{km/s/Mpc}$ for the flat $\Lambda$CDM model;
$w_0 = -1.22^{+0.18}_{-0.15}$ for the flat $w$CDM model; and $w_{a} = -1.12^{+0.45}_{-0.83}$  for the flat Chevallier-Polarski-Linder model at the 1$\sigma$ confidence level.
Our results with the GOLD and FULL sample are almost identical, which are more stringent than the previous results with GRBs.
\end{abstract}
\begin{keywords}
	gamma-rays: general < Resolved and unresolved sources as a function of wavelength, cosmology: observations < Cosmology
	
\end{keywords}

\section[INTRODUCTION]{INTRODUCTION}\label{sec: Intro}
Gamma-ray bursts (GRBs) are the most intense explosions observed to occur in the  high redshift range measured up to $z=8\sim10$, e.g., GRB 090423 at $z=8.2$ \citep{Tanvir2009, Salvaterra2009}, and GRB 090429 at of $z=9.4$ \citep{Cucchiara2011}, which can be used to probe the universe beyond the type Ia supernovae (SNe Ia)  observed  at the  maximum redshift about $z\sim2$. 
Recently, several luminosity (or energy) relations of GRBs  
between measurable spectroscopic properties that can be related to an intrinsic burst physical property and their luminosity or energy have been proposed, e.g. the prompt emission correlations \citep{Norris_2000, fenimore_2000, Amati_2002, Schaefer2003, Yonetoku_2004, Ghirlanda_2004}, and prompt-afterglow emission correlations  \citep{Liang_2005, Dainotti_2008, Dainotti_2011, Dainotti_2016b, Dainotti_2017, Dainotti_2018, Bernardini_2012, Izzo_2015, Tang_2019}. 
Therefore, GRBs can be used as standard candles to probe cosmology at high redshift range e.g., \cite{Dai2004, Liang2006, Wang2006, Ghirlanda2006, Schaefer2007} and \cite{Amati2013}.
See \cite{luongo2021roadmap} for reviews.

In order to avoid the circularity problem \citep{Ghirlanda2006}, the simultaneous method \citep{Amati_2008} in which the parameters of the relationship and the cosmological model fitting simultaneously has also been proposed.
\cite{Liang2008} proposed a cosmological model-independent method to calibrate GRBs relations at low redshift  from SNe Ia  without any cosmological assumption. Therefore, the GRB Hubble diagram at high redshift can be used to constrain  cosmological models \citep{Wei2009, Wei2010, Liang2010, Liang2011, Liu2015}.
On the other hand,  \cite{Amati_2019} proposed an alternative method to calibrate the GRB relation  of 193 GRBs 
by using the observational Hubble data (OHD) obtained with the cosmic chronometers (CC) method 
\citep{Montiel2021, Luongo2021b, Luongo2023}.

The so-called Amati relation \citep{Amati_2002}, which connects the spectral peak energy ($E_{\rm p}$) and the isotropic equivalent radiated energy ($E_{\rm iso}$) has been widely used to estimate cosmological parameters \citep{Amati_2006, Amati_2008, Amati2009, Amati2013, Amati_2019, Wei2010, Wang2016, Demianski_2017a, Demianski_2017b, Lusso_2019}.
Quantified by the $T_{90}$ parameter (the time interval over which $90\%$ of the gamma-ray emission is measured) with a division at 2s by the gamma-ray duration,  GRBs can be classified into 
the \emph{short} GRBs (SGRBs) and the \emph{long} GRBs (LGRBs), which are mostly associated to the merger of compact objects (Type I) and associated to the collapse of certain types of massive stars  (Type II), respectively.
Previous analyses of the $E_{\rm p}-E_{\rm iso}$ plane of GRBs shows that different classes of GRBs exhibit different behaviours: while normal long GRBs follow the $E_{\rm p} - E_{\rm iso}$ correlation, short GRBs do not (e.g., \cite{Amati_2006}).

On a mission to study the universe at high energies, the \emph{Fermi} Gamma-ray Space Telescope was launched on 2008 June, including two main instruments: the Gamma-ray Burst Monitor (GBM) and the Large Area Telescope (LAT) with unprecedented sensitivity to gamma-rays overlapping the GBM energy range and extending it up to 300 GeV \citep{von2020fourth} which provides a powerful tool for studying GRBs.
Recently, \cite{Amati2009} found that \emph{Fermi} GRBs are fully consistent with the Amati relation, based on Fermi/GBM and LAT spectral measurements (with a cutoff at GRB090424);
\cite{Wei2010} used the cosmology-independent method \citep{Liang2008} to calibrate the Amati relation with 109 GRBs, including 14 additional GRBs (with a cutoff at GRB091208B detected by \emph{Fermi});
\cite{Wang2016} standardized the Amati relation by the cosmology-independent calibration method and the simultaneous method with 151 GRB data, including further 42 GRBs (with a cutoff at GRB140213A detected by \emph{Fermi}).
\cite{Dirirsa2019} found that the Amati relation is satisfied by the 25 long GRBs with Fermi-LAT sample (from June 2008 to September 2017) detected simultaneously by \emph{Fermi} GBM and LAT, together with a sample of 94 GRBs selected from \cite{Wang2016}, which are not in Fermi sample or counted twice;
\cite{Khadka2021} compiled  a total 220 GRB sample (the A220 sample) in which a data set of 118 GRBs (the A118 sample)\footnote{A short GRB090510 discarded from the 119 GRBs \citep{Dirirsa2019,Khadka2020}.} with the smallest intrinsic dispersion is  suitable for constraining cosmological parameters. 

More recently, \cite{Montiel2021} obtained a refined set of 74 \emph{long} GRBs from a total 107 data
of \emph{Fermi}-GBM catalogue (from August 2008 to March 2019) at $0.117\le z\le5.283$ based on criteria such as redshift and spectral signal-to-noise ratio, to calibrate the Amati relation through a Bezier parametric curve fitted from the OHD at $z\le1.43$.
\cite{Jia_2022} tested the Amati relation by using 221 GRBs (the J221 sample) which is based on the previous data from \citet{Wang2016,Amati_2019} that appear in both \emph{Swift} and \emph{Fermi} catalogs, and 49 GRBs from \emph{Fermi} catalog (from 2013 to March 2021).
\cite{Dainotti2023} compiled a total sample of 86 GRBs with the relationship between the spectral and temporal indices using closure relation from the GRBs that have been fitted in the Fermi-LAT Second GRB Catalog (2FLGC) with 186 GRBs observed from 2008 to 2018. 

With the enhancement of the number of GRBs with measured redshift and spectral parameters,
\cite{Liang2022} used a Gaussian Process to  calibrate the A219 sample \footnote{Removed GRB051109A, which are counted twice in the A220 sample \citep{Khadka2021}. For details, see \cite{Liang2022}.}  from the Pantheon sample \citep{Scolnic2018} which contains 1048 SNe  and constrained Dark Energy models with GRBs at high redshift. 
\cite{LZL2023} calibrated GRBs from the latest 32 OHD using Gaussian Process to construct the GRB Hubble diagram.
\cite{Xie2023} calibrated the J221 sample  with the Amati relation by the Gaussian Process from the Pantheon+ sample \citep{Scolnic2022} which contains 1701 SNe light curves of 1550 spectroscopically confirmed SNe Ia.
\cite{Zhang2023} calibrated the Amati relation by the machine learning methods for reconstructing distance-redshift relation from the Pantheon+ sample.
\cite{Wang2024} constrain the emergent dark energy models with the cosmology-independent GRBs at high-redshift and OHD to find a large value of $H_0$ which is close to the results of local measurement from the  \emph{Supernova H0 for the Equation of State} (SH0ES) Collaboration \citep{Riess2022}.

In this paper, we present a sample of long GRBs  from 15 years of the Fermi-GBM catalogue with redshift measured at $0.0785\le z\le8.2$. The calibration is performed following the recent work \citep{Li_2023} with a compilation of OHD 
without assuming an a priori cosmological model. 
Our sample originates from a single catalogue, which are carefully selected to avoid large errors in the determination of luminosity parameters of the Amati relation, thus avoiding selection biases and other instrument-associated systematics.
We calibrate the Amati relation with the Fermi-GBM catalogue at low-redshift from the latest OHD  with the cosmic chronometers method by using a Gaussian Process to obtained GRBs at high-redshift, which can be used to constrain cosmological models in cosmology-independent way.

The paper is organized as follows. In Sec. \ref{sec:dSample}, we present in detail our \emph{Fermi}-GBM LGRBs sample. In Sec. \ref{sec:calAmati}, we reconstruct OHD using GaPP and calibrate the Amati relation. In Sec. \ref{sec:conDEModel},  we present the datasets included in our suite of observations to fit parameters of DE models. We discuss our results and draw conclusion in Sec. \ref{sec:conclusion}.

\section[Data Sample]{Data Sample}\label{sec:dSample}
The GBM onboard the \emph{Fermi} Gamma-ray Space Telescope is made up of 
12 sodium iodide (NaI) detectors 
that cover the energy range 8 keV to $\sim1$ MeV, and two bismuth germanate oxide (BGO) detectors (\texttt{b0} and \texttt{b1})
sensitive from 200 keV to 40 MeV.
The axes of the NaI detectors in four groups of three on the corners of the spacecraft are oriented to optimize all-sky coverage and enable the localization of GRBs by comparing the relative observed source rates in each detector. The BGO detectors are located on opposite sides of the spacecraft to enable an all-sky view \citep{Meegan_2009}.
The onboard GBM trigger system for detecting GRBs was first enabled on 2008 July 12 \citep{Paciesas2012}.

The \emph{Fermi}-GBM science team releases GRB catalogs on a regular basis that list the main characteristics of triggered bursts, compiling the data of several completed mission years, including the first catalog: the first two years \citep{Paciesas2012},  second: the four years \citep{von2014second}, third: the six years \citep{Bhat_2016}, and four: the 10 years \citep{von2020fourth}.
The spectral catalogs were accompanied by \citet{Goldstein2012} and  \citet{gruber2014fermi} for the first two and four mission years.
\cite{Poolakkil2021} have updated the 10 year spectral catalog (from June 2008 to March 2018) including 122 long GRBs with redshift.
\cite{Montiel2021} obtained a refined set of 74 \emph{long} GRBs from a total 107 data
of \emph{Fermi}-GBM catalogue (from August 2008 to March 2019) at $0.117\le z\le5.283$.

In this work, we collect 15 years (2008-2023) of GRB data observed  by \emph{Fermi}-GBM with redshift measured.
All the GRB data are obtained from the FERMIGBRST Catalog\footnote{\url{https://heasarc.gsfc.nasa.gov/W3Browse/fermi/fermigbrst.html}}, hosted on NASA High Energy Astrophysics Science Archive Research Center (HEASARC),
which contains all GBM triggered events classified as GRBs.
For the case of GRBs which present no value for the spectral parameters after 2018 June, we use the preliminary spectral analysis results with best fit by a BAND function\footnote{For a very wide energy band of Fermi-GBM, the GRBs prompt emission spectrum $\Phi(E)$ can be usually best fitted by the BAND model
	as the spectral parameters (the observed peak energy $E_{\rm p}$, the low and high energy spectral index: $\alpha$ and $\beta$) \citep{ZHZ2023}, 
	which is an empirical spectral function with a broken power law \citep{Band_1993}:  
	\begin{equation}
		\label{equ:band}
		\Phi(E) =A\begin{cases}
			(\frac{E}{\rm 100keV})^{\alpha}e^{-(2+\alpha )\tfrac{E}{E_{\rm p}}},~~~~ \text{if } E \le \frac{\alpha-\beta}{2+\alpha}E_{\rm p}
			\\
			\\
			(\frac{E}{\rm 100keV})^{\beta}(\frac{\alpha-\beta}{2+\alpha}E_{\rm p})^{(\alpha-\beta)}e^{(\beta-\alpha )}, ~\rm otherwise.
		\end{cases}
	\end{equation}
	where 
	the normalization parameter $A$ is chosen to ensure continuity at the break and to match the observed burst brightness.
In this work, the BAND function which has been used in many  previous analysis \citep{Wang2016,Demianski_2017a,Amati_2019,Montiel2021,Jia_2022} is chosen as the reference model for simplicity.
It should be noted that other spectral models (COMP: An exponentially attenuated power law, namely "comptonized", SBPL: A smoothly broken power law and PLAW: A single power law) \citep{Poolakkil2021} can be chose to fit the spectra of GRBs.
\cite{Poolakkil2021} have used 10 years of GBM data and the known redshift for $\sim$130 GRBs  to obtain different rest-frame energetics with the BAND model and the COMP model, respectively.}
for 19 GRBs\footnote{GRB180728A, GRB181020A, GRB190114C, GRB190324A, GRB190829A, GRB200524A, GRB200613A, GRB200829A, GRB201020B, GRBGRB201216C, GRB210204A, GRB210619B, GRB211023A, GRB220101A, GRB220107A, GRB220527A, GRB220627A, GRB230204B, GRB230812B.} reported in GCN (The Gamma-ray Coordination Network) Circulars Archive\footnote{\url{https://gcn.nasa.gov/circulars}}.
For the GRBs of the preliminary spectral analysis results with best fit NOT a BAND function, we download the GBM Burst Data from FTP archive\footnote{\url{https://heasarc.gsfc.nasa.gov/FTP/fermi/data/gbm/bursts/}} to calculate parameters  with best fit by BAND model by  employing
the Gamma-Ray Spectral Fitting Package (RMFIT V4.3.2)\footnote{\url{https://fermi.gsfc.nasa.gov/ssc/data/p7rep/analysis/rmfit/}} for 14 GRBs\footnote{GRB181010A, GRB190613A, GRB190719C, GRB191011A, GRB201020A, GRB201021C, GRB210104A, GRB210610A, GRB210610B, GRB210722A, GRB210731A, GRB220521A, GRB221226B, GRB230818A.}.
The spectroscopic or photometric redshifts of GRBs are obtained from the \emph{Swift} database\footnote{\url{https://swift.gsfc.nasa.gov/archive/grb\_table.html}}  
and the webpage of J. Greiner\footnote{\url{http://www.mpe.mpg.de/~jcg/grbgen.html}}, which compiles 
and published papers.

We identify a total 187 GRBs with redshift measured (with a cutoff at GRB230818).
In order to obtain suitable long GRB data with redshifts, we excluded some GRBs based on the following criteria:

\begin{enumerate}
	\item Short GRBs: There are 26 short GRBs\footnote{It is worth noting that 2 GRBs
		with $T_{\rm 90}$ > 2s should be  short GRBs: GRB170817A (GBM trigger 170817529 with $T_{\rm 90}$ = 2.048)\citep{Goldstein_2017} and GRB 180618A (GBM trigger 180618030 with $T_{\rm 90}$ = 3.712)\citep{Jordana2022}.} from the identified 187 GRBs with redshift.
	\item Unsafe Redshifts 
	: There are 2 GRBs\footnote{For GRB100802A, the data are not sufficient to estimate a definite photometric redshift, but place an upper limit redshift of $z<3.1$ \citep{Perley_2016}. For GRB190530A, any significant absorption or emission lines in the low-resolution spectrum is not detected, but an upper limit of $z<2.2$ has been inferred based on the detected continuum \citep{Heintz_2019}.} with only limits and one GRB\footnote{For GRB110721A, the redshift is not known for spectroscopy of the counterpart suggested two possible redshifts, $z = 0.382$ and $3.512$ 
		\citep{Iyyani_2013}.  
	}with two possible redshifts.
	\item Significant uncertainties in the spectral parameters:
	There are 7 GRBs\footnote{The relative error of $E_{\rm peak}$ of GRB080928,  GRB100728B, GRB110128A, GRB120729, GRB150120, GRB160620, and GRB221009 are significant large.} with significant uncertainties of $E_{\rm p}$ which only present poor contribution to fitting procedure.

	By discarding the above short GRBs, GRBs with unsafe redshifts and significant uncertainties in the spectral parameters, we can obtain 151 long GRBs with identified redshift at $0.0785 \leq z \leq 8.2$ to be considered as the FULL sample. 
	More careful selections should be required  to reduce the uncertainties of GRB data and redshifts measured.
	\item Photometric Redshifts: There are 8 GRBs\footnote{GRB080916, GRB120922, GRB131229, GRB151111, GRB151229, GRB161001, GRB180418, GRB200829.} with redshift obtained through photometric method, relying on photometric information rather than spectral features, leading to differences between measured values and actual redshifts.
	\item Uncertain Redshifts: There are 2 GRBs\footnote{For GRB140713A, \cite{Higgins2019} obtained $z = 0.935$ with the emission spectrum of the likely host galaxy; however, \cite{Schroeder2022} obtained $z = 0.935\pm0.002$ with identified [O II] and [O III]. For GRB190613A, the redshift  due to Lyman alpha  at $z = 2.78$ is tentative currently due to the lack of corroborating features  \citep{Cunningham_2019}.}  with uncertain redshifts.
	\item The peak flux selection: We also considered the impact of peak flux selection with a peak flux threshold in the trigger energy range that is
	typically $50\%$ higher than the trigger threshold \citep{Atteia2017}. 
	There are 22 GRBs with a 1s peak flux in the energy range ($50-300 \rm keV$) lower than 1.5 times the detection threshold of $P=0.70 \rm ph~cm^{-2}~s^{-1}$, in which 18 GRBs are not counted twice.
	
\end{enumerate}

\renewcommand{\arraystretch}{1.2}
\begin{table*}\footnotesize
	\caption{Burst Not Included in  the FULL sample and the GOLD sample.}
	\label{tbl:disGRB}
	\begin{threeparttable}
		\begin{tabular}{|lll|lll|lll|}
			\hline\hline
			\textbf{GRB} & \textbf{Cause} & \textbf{Note} & \textbf{GRB} & \textbf{Cause} & \textbf{Note} & \textbf{GRB} & \textbf{Cause} & \textbf{Note}\\ \hline
			080905A & short		        & $T_{90}$=0.96               & 120729A & SU             & $E_{\rm p,obs}^{*} = 6.614$ & 170817A & short 	      & $T_{90}$=2.048\\
			080905B	& low-flux\tnote{$a$}& $P = 0.5815$               & 120922A & photometric-$z$& $z$=3.1(ph)                 & 171222A & low-flux       & $P = 0.1903$\\
			080916C & photometric-$z$   & $z$=4.35(ph)\tnote{$b$}     & 121211A & low-flux       & $P = 0.5809$                & 180418A & photometric-$z$& $z$=1.55(phh)\\
			080928	& SU\tnote{$c$}     & $E_{\rm p,obs}^{*} = 3.395$ & 130515A & short 	     & $T_{90}$=0.256              & 180618A & short 	      & $T_{90}$=3.712\\
			081109A	& low-flux          & $P = 0.7858$                & 130612A & low-flux       & $P = 0.2384$                & 180727A & short 	      & $T_{90}$=0.896\\
			090423	& low-flux          & $P = 0.5781$                & 130716A & short 	     & $T_{90}$=0.768              & 180805B & short 	      & $T_{90}$=0.96\\
			090510  & short 	    	& $T_{90}$=0.96               & 131004A & short 	     & $T_{90}$=1.152              & 181010A & low-flux       & $P = 0.0541$\\
			090927  & short 	    	& $T_{90}$=0.512              & 131229A & photometric-$z$& $z$=1.04(phh)\tnote{$b$}    & 190530A & unsafe-$z$     & $z$ < 2.2\\
			100117A & short 	    	& $T_{90}$=0.256              & 140304A & low-flux       & $P = 0.9886$                & 190613A & uncertain-$z$  & $z$=2.78(?)\\
			100206A & short 	    	& $T_{90}$=0.176              & 140623A & low-flux       & $P = 0.5639$                & 191011A & low-flux       & $P = 0.7477$\\
			100625A & short 	    	& $T_{90}$=0.24               & 140713A & uncertain-$z$  & $z = 0.935\pm0.002$         & 191031D & short 	      & $T_{90}$=0.256\\
			100728B	& SU            	& $E_{\rm p,obs}^{*} = 2993$  & 150101B & short 	     & $T_{90}$=0.08               & 200219A & short 	      & $T_{90}$=1.152\\
			100802A & unsafe-$z$ 	    & $z$ < 3.1 \tnote{$d$}       & 150120A & SU		     & $E_{\rm p,obs}^{*} = 16.73$ & 200411A & short 	      & $T_{90}$=1.44\\
			101219B	& low-flux          & $P = 0.6811$                & 151111A & photometric-$z$& $z$=3.5(ph)                 & 200826A & short 	      & $T_{90}$=1.136\\
			101224A & short 	    	& $T_{90}$=1.728              & 151229A & photometric-$z$& $z$=1.4(phh)                & 200829A & photometric-$z$& $z$=1.25(ph)\\
			110106B	& low-flux          & $P = 0.8208$                & 160408A & short 	     & $T_{90}$=1.056              & 201020A & low-flux       & $P = 0.612$\\
			110128A	& SU				& $E_{\rm p,obs}^{*} = 139.7$ & 160623A & SU		     & $E_{\rm p,obs}^{*} = 4.632$ & 201021C & low-flux       & $P = 0.549$\\
			110721A & unsafe-$z$ 	    & $z$=3.512(?)\tnote{$d$}     & 160624A & short 	     & $T_{90}$=0.384              & 201221D & short 	      & $T_{90}$=0.144\\
			110818A	& low-flux          & $P = 0.7446$                & 160821B & short 	     & $T_{90}$=1.088              & 210323A & short 	      & $T_{90}$=0.96\\
			111107A	& low-flux          & $P = 0.563$                 & 161001A & photometric-$z$& $z$=0.67(phh)               & 221009A & SU             & $E_{\rm p,obs}^{*} = 2.35$\\
			111117A & short 	    	& $T_{90}$=0.432              & 170113A & low-flux       & $P = 0.5287$                & & &\\
			120118B	& low-flux          & $P = 0.7201$                & 170127B & short          & $T_{90}$=1.728              & & &\\
			\hline
		\end{tabular}
		\begin{tablenotes}
			\footnotesize
			\item[$a$] low-flux: The 1s peak flux in the energy range ($50-300 \rm keV$) lower than 1.5 times the detection threshold of $P=0.70 \rm ph~cm^{-2}~s^{-1}$.
			\item[$b$] SU: Significant uncertainties in the spectral parameters. The relative error of $E_{\rm p}$ ($E_{\rm p,obs}^{*}$) is significant large.
			\item[$c$] ph=photometric redshift; phh=photometric redshift of host galaxy;
			\item[$d$]"<"=upper limit redshift; "?"=controversial redshift;
		\end{tablenotes}
	\end{threeparttable}
\end{table*}

\begin{figure}
	\centering
	\includegraphics[width=\columnwidth]{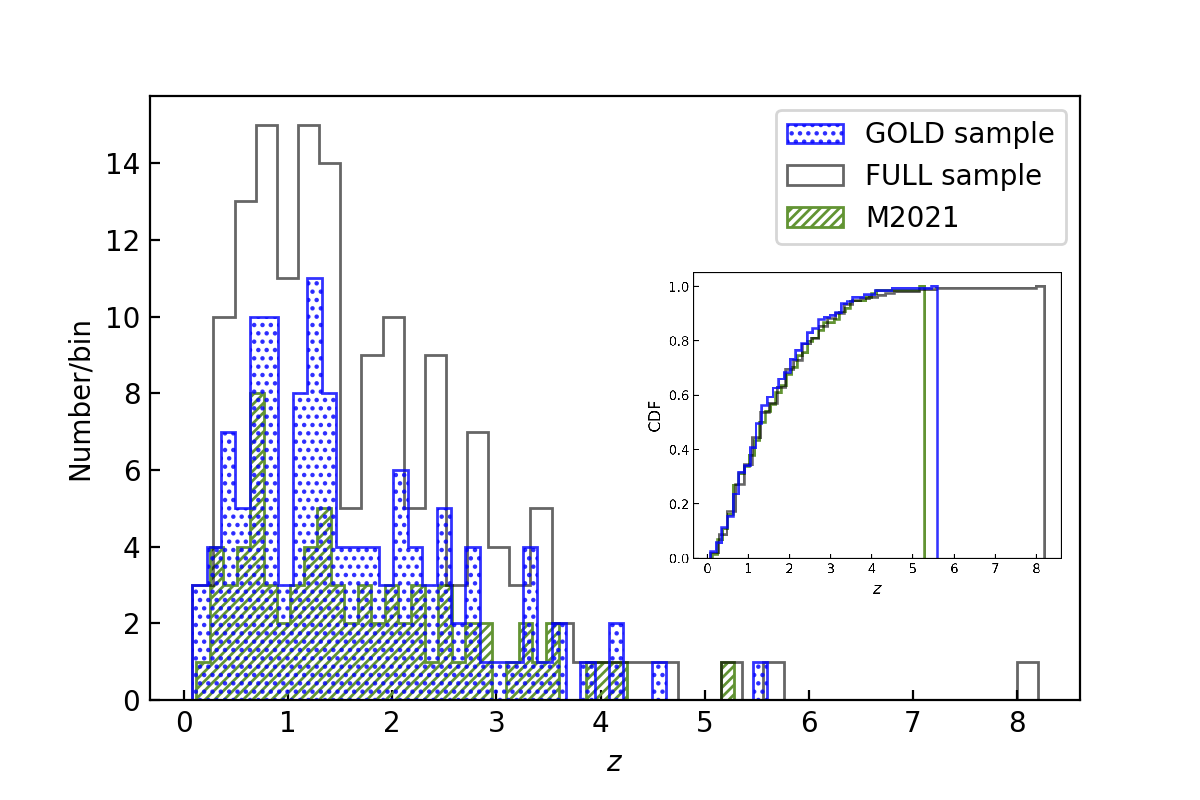}
	\caption{The redshift distribution for the Fermi-GBM GRBs with the FULL sample (\emph{black histogram}) of 151 GRBs at $0.0785 \leq z \leq 8.2$, the GOLD sample (\emph{blue  histogram}) of 123 GRBs at $0.0785 \leq z \leq 5.6$, 
		and the M2021 sample (\emph{green  histogram}) of 74 GRBs at $0.117\le z\le5.283$.
		The cumulative distribution function (CDF) of the FULL sample (\emph{black curve}), the GOLD sample (\emph{blue curve}),
		and the M2021 sample (\emph{green curve}) are indicated in the inset, respectively.}
	\label{fig:redshifts}
\end{figure}

If further discarding GRBs with photometric redshifts and uncertain redshifts, and GRBs lower than the peak flux selection, we can get a \emph{Fermi}-GBM GRB sample consists of 123 long GRBs to be considered as the GOLD sample from the FULL sample of 151 long GRBs with identified redshift, which covers from GRB 080804 to GRB 230818A at the redshift range $0.0785 \leq z \leq 5.6$.
The bursts not included in the  GOLD and FULL sample are shown in Table \ref{tbl:disGRB}.
The redshift distribution for the GOLD and FULL samples are shown in Figure \ref{fig:redshifts}.
For comparison, 
the M2021 sample \citep{Montiel2021} of 74 GRBs  at $0.117\le z\le5.283$  are also shown in Figure \ref{fig:redshifts}.
In Appendix, we list the \emph{Fermi}-GBM GRBs with the FULL and GOLD samples together with the redshifts ($z$), the $T_{90}$ parameter;  as well as the spectral parameters ($E_{\rm p}$,  $\alpha$, $\beta$) and the measured bolometric fluence $S_{\rm bolo}$ with the associated errors in Table \ref{tab:LGRBs} and Table \ref{tab:addLGRBs}.
\footnote{$S_{\rm bolo}$ can be computed from the observed fluence $S$ multiplying by the \emph{k}-value in the standard rest-frame energy band $\rm 1 - 10^4 keV$ \citep{Poolakkil2021},
	\begin{equation}
		\label{equ:Sbolo}
		k = \frac{ \int_{1/(1+z)}^{10^4/(1+z)} E\Phi(E) dE}{ \int_{E_{\rm min}}^{E_{\rm max}} E\Phi(E) dE}
	\end{equation}
	here the detection limits of GBM: $E_{\rm min} = 10$ keV and $E_{\rm max} = 1000$ keV.
	The error of $k$ can be calculate by using the following error propagation formula:
	\begin{equation}
		\label{equ:eSbolo}
		\begin{aligned}
			(\sigma_{k})^2& =(\frac{\partial k}{\partial\alpha})^2(\sigma_{\alpha})^2 +(\frac{\partial k}{\partial\beta})^2(\sigma_{\beta})^2
			+(\frac{\partial k}{\partial A})^2(\sigma_{A})^2+(\frac{\partial k}{\partial E_{\mathrm{p}}})^2(\sigma_{E_{p}})^2
		\end{aligned}
	\end{equation}
}

\section[Calibration]{Calibration of Amati Relation}\label{sec:calAmati}

The Amati relation relates the spectral peak energy ($E_{\rm p,i}$)  and the isotropic energy ($E_{\rm iso}$), which can be expressed in the functional form,
\begin{equation}
	\label{equ:Amati}
	\log_{10}(\frac{E_{\rm iso}}{\rm erg}) = a + b\log_{10}(\frac{E_{\rm p,i}}{\rm 300keV})
\end{equation}
where $a$ and $b$ are free parameters to be determined from the data, the spectral peak energy in the GRB cosmological rest frame $E_{\rm p,i} = E_{\rm p}(1+z)$, and $E_{\rm iso}$
\begin{equation}
	\label{equ:Eiso}
	E_{\rm iso} = 4\pi d_{\rm L}^2S_{\rm bolo}(1 + z)^{-1}
\end{equation}
The luminosity distance $d_{\rm L}$, for a flat spatially universe, can be calculated by 
\begin{equation}
	\label{equ:dlohd}
	d_{\rm L}^{\rm cal} = c(1+z)\int_{0}^{z} \frac{dz^{'}}{H(z^{'})}.
\end{equation}

We use a Gaussian process to reconstruct $H(z)$ at redshift $z$ with the observational data. 
Gaussian process \citep{seikel_2012} is a nonparametric technique, in which the reconstruct only depends on the chosen kernel function and the data observed. In the Gaussian process, the reconstructed function $\boldsymbol{f_*}$ from  the data $\boldsymbol{y}$ which is known from observations can be described with its mean function and covariance function:
\begin{equation}\small
	\begin{aligned}
		&\bar{\boldsymbol{f}_*}=\boldsymbol{\mu}_* + K(\boldsymbol{Z_*,Z})\left[K(\boldsymbol{Z,Z}) + \mathbf{C}\right]^{-1}(\boldsymbol{y} - \boldsymbol{\mu})
		\\
		&\mathrm{cov}(\boldsymbol{f}_*)=K(\boldsymbol{Z_*,Z_*})-K(\boldsymbol{Z_*,Z})\left[K(\boldsymbol{Z,Z})+\mathbf{C}\right]^{-1}K(\boldsymbol{Z,Z_*})
	\end{aligned}
\end{equation}
where $\boldsymbol{\mu}_*$ is the priori assumed mean of $\boldsymbol{f}_*$.
For a set of input points $\boldsymbol{Z} = \{z_i\}$, the covariance matrix $K\boldsymbol{(Z, Z)}$ is given by $[K\boldsymbol{(Z, Z)}]_{ij} = \boldsymbol{k}(z_i, z_j)$. In this work, we use the squared exponential covariance function\citep{seikel_2012}: $\boldsymbol{k}(z,\widetilde{z}) = \sigma^{2}_{f}\exp(-\frac{(z - \widetilde{z})^2}{2l^2})$, with the two hyperparameters $l$ and $\sigma_{f}$ which can be optimized and determined by maximizing the log marginal likelihood.

\begin{figure}
	\centering
	\includegraphics[width=\columnwidth]{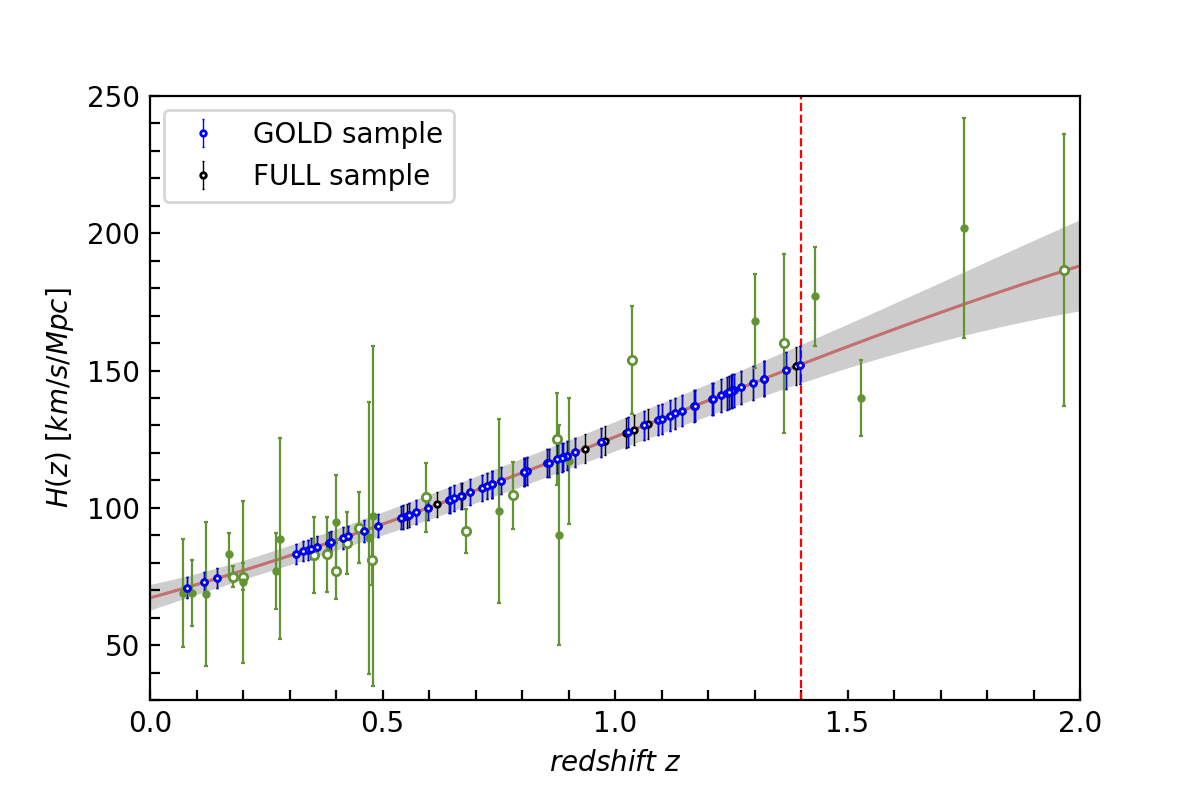}
	\caption{The reconstructed results from OHD through Gaussian process.
		The green cycles and dots indicate  the 15 OHD with the full covariance matrices and the other 17 OHD without covariance, respectively.
		The red curve presents the reconstructed function with the $1\sigma$ uncertainty from  OHD by taking into account the full covariance matrices.
		The GRBs with the FULL sample (\emph{black cycles})  and the GOLD sample (\emph{blue cycles}) at $z < 1.4$ are reconstructed from OHD through the Gaussian process. The red dotted line denotes $z = 1.4$.}
	\label{fig:fHz}
\end{figure}

We use the public python package GaPP\footnote{\url{https://github.com/astrobengaly/GaPP}} with the observational data.
The latest 32 OHD  in the range $0.07<z<1.965$ (see \cite{Moresco_2022, Li_2023} and reference therein) are used to reconstruct $H(z)$. The total covariance matrix\footnote{\url{https://gitlab.com/mmoresco/CCcovariance/-/tree/master?ref_type=heads}} as the combination of the statistical and systematic part \citep{Moresco_2020} of 15 $H(z)$ estimates  from the measurements \citep{Moresco_2012, Moresco_2016, Moresco_2015} in the range $0.179<z<1.965$  
has to be taken into account. 
In order to calibrate the Amati relation,
we choose $z = 1.4$ \citep{Liang2022,Li_2023} to divide GRBs into low- and high-redshift sample. 
The reconstructed results through Gaussian process with the $1 \sigma$ uncertainty from OHD are plotted in Figure \ref{fig:fHz}.

The Markov Chain Monte Carlo (MCMC) of Python package emcee \citep{Foreman-Mackey_2013} is used to calibrate the Amati relation
at $z < 1.4$ by using the fitting methods with the likelihood function proposed by \cite{dagostini2005fits}\footnote{The likelihood function proposed by \cite{dagostini2005fits} is: $\mathcal{L_{\rm D}}\propto \prod_{i=1}^{N_{1}} \frac{1}{\sigma} \times \exp\left [ -\frac{\left [ y_{i}  - y(x_i,z_i;a,b)\right ]^2 }{2(\sigma_{\rm int}^2+\sigma^2_{y,i}+b^2\sigma^2_{x,i})}  \right ]$,
	with
	$\sigma_{\rm int}$ means the intrinsic scatter.} and \citet{reichart2001dust}\footnote{The likelihood function proposed by \citet{reichart2001dust}  can be written as \citep{Lin_2015,Li_2023}:
	$\mathcal{L_{\rm R}}\propto \prod_{i=1}^{N_{1}} \frac{\sqrt{1+b^2}}{\sigma} \times \exp\left [ -\frac{\left [ y_{i}  - y(x_i,z_i;a,b)\right ]^2 }{2(\sigma_{\rm int}^2+\sigma^2_{y,i}+b^2\sigma^2_{x,i})}  \right ]$,
	here $\sigma_{\rm int} = \sqrt{\sigma^2_{y,\rm int} + b^2\sigma^2_{x,\rm int}}$, in which $\sigma^2_{y,\rm int}$ and $\sigma^2_{x,\rm int}$ are the intrinsic scatter along the axes.} which has the advantage of not requiring the arbitrary choice of an independent variable in $x$-axis and $y$-axis.

\begin{figure}
	\centering
	\includegraphics[width=\columnwidth]{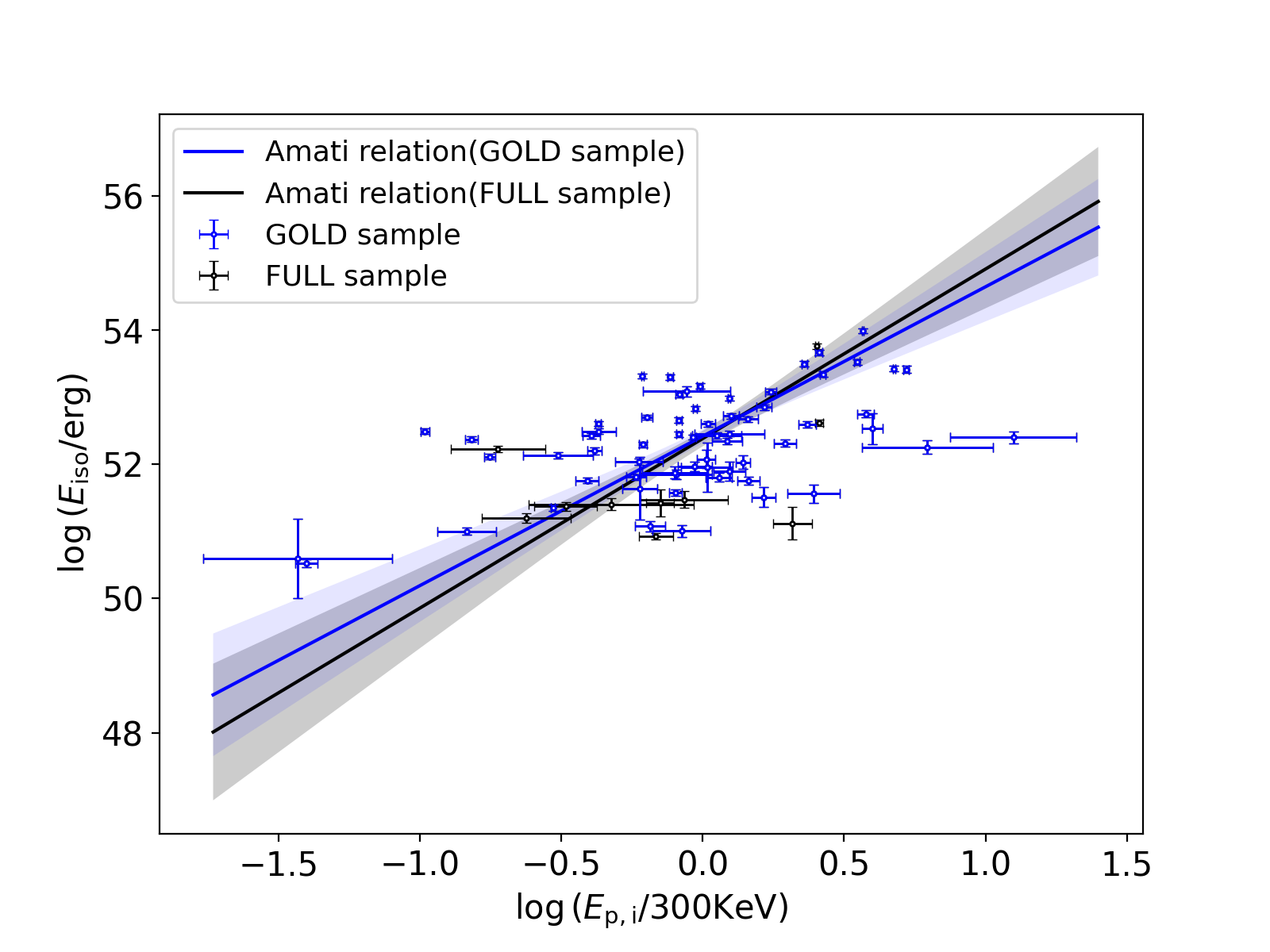}
	\caption{The best-fit calibration of Amati relation by BCES-orthogonal method with the FULL sample (\emph{black cycles}) and the GOLD sample (\emph{blue cycles}) at $z < 1.4$. For the FULL sample, the Spearman$'$s rank coefficient of the correlation is $\rho$ = 0.49, and the P-value = $6.9\times10^{-6}$. For the GOLD sample, the Spearman$'$s rank coefficient of the correlation is $\rho$ = 0.45, and the P-value = $1.82\times10^{-4}$.}
	\label{fig:bces}
\end{figure}

We also take into account the possible intrinsic scatter of the data with respect to the linear fit and the measurement errors in both variables and employ the Python module \citet{nemmen_universal_2012}\footnote{\url{https://github.com/rsnemmen/BCES}} by the Bivariate Correlated Errors and intrinsic Scatter (BCES) method \citep{akritas_linear_1996}.
The calibration results of the Amati relation by the three methods are summarized in Table \ref{tab:amati}.
We find that the fitting results by the likelihood function \citep{dagostini2005fits} are consistent with the ones by BCES($y|x$);
and the fitting results by the likelihood function \citep{reichart2001dust} are consistent with the ones by BCES-orthogonal\footnote{The BCES-Orthogonal least squares: line that minimizes orthogonal distances, which should be used when it is not clear which variable should be treated as the independent one, whereas the BCES-bisector approach which bisects the $y|x$ and $x|y$ is self-inconsistent \citep{Hogg_2010}.}.
We plot the Amati relation with the FULL and GOLD samples at $z < 1.4$ by the BCES-orthogonal method in  Figure \ref{fig:bces}. The results of the intercept $a$ and the slope $b$ with the FULL and GOLD sample are well consistent at $1\sigma$, while the difference of the slope $b$ between the FULL and GOLD sample is apparently significant, in which the value of the slope $b$ with the GOLD sample is close to the theoretical value of the slope with the Amati relation: 2.
Our results at $z < 1.4$ by the BCES-orthogonal method are consistent with \citet{Montiel2021},
which calibrated the Amati relation with the OHD at $z<1.43$.

\begin{table}\scriptsize
	\caption{Calibration results (the intercept $a$, the slope $b$, the intrinsic scatter $\sigma_{int}$, and the covariance matrix by BCES methods) of the Amati relation with  the FULL sample and the GOLD sample at $z < 1.4$ by D'Agostini method and the Reichart method\citep{reichart2001dust,dagostini2005fits} and the BCES methods \citep{akritas_linear_1996} with BCES($y|x$) and BCES-Orthogonal least squares.}
	\label{tab:amati}
	\centering
	\begin{threeparttable}
		\renewcommand\arraystretch{1.5}
		\begin{tabular}{lcccc}
			\hline\hline
			Sample/Methods & \boldmath{$a$} & \boldmath{$b$} & \boldmath{$\sigma_{int}$}/\textbf{cov} \\ \hline
			FULL/D'Agostini & $52.35\pm{0.08}$ & $0.88\pm{0.18}$ & $0.65^{+0.05}_{-0.06}$ \\ \hline
			GOLD/D'Agostini & $52.41\pm{0.08}$ & $0.83\pm{0.18}$ & $0.61^{+0.05}_{-0.07}$ \\ \hline
			FULL/Reichart & $52.42\pm{0.11}$ & $2.52^{+0.40}_{-0.19}$ & $0.50\pm{0.29}$ \\ \hline
			GOLD/Reichart & $52.46\pm{0.12}$ & $2.36^{+0.38}_{-0.33}$ & $0.50\pm{0.29}$ \\ \hline
			FULL/BCES($y|x$) & $52.33\pm{0.08}$ & $0.92\pm{0.18}$ &
			$\begin{pmatrix}
				0.006 & 0.003\\
				0.003 & 0.033
			\end{pmatrix}$
			\\
			\hline
			GOLD/BCES($y|x$) & $52.39\pm{0.08}$ & $0.86\pm{0.18}$ &
			$\begin{pmatrix}
				0.006 & 0.002\\
				0.002 & 0.032
			\end{pmatrix}$
			\\
			\hline
			FULL/BCES-Ort & $52.39\pm{0.11}$ & $2.46\pm{0.58}$ &
			$\begin{pmatrix}
				0.013 & -0.003\\
				-0.003 & 0.331
			\end{pmatrix}$
			\\
			\hline
			GOLD/BCES-Ort & $52.43\pm{0.11}$ & $2.16\pm{0.57}$ &
			$\begin{pmatrix}
				0.012 & -0.007\\
				-0.007 & 0.324
			\end{pmatrix}$
			\\
			\hline
			\hline
		\end{tabular}
	\end{threeparttable}
\end{table}

\begin{figure}
	\centering
	\includegraphics[width=\columnwidth]{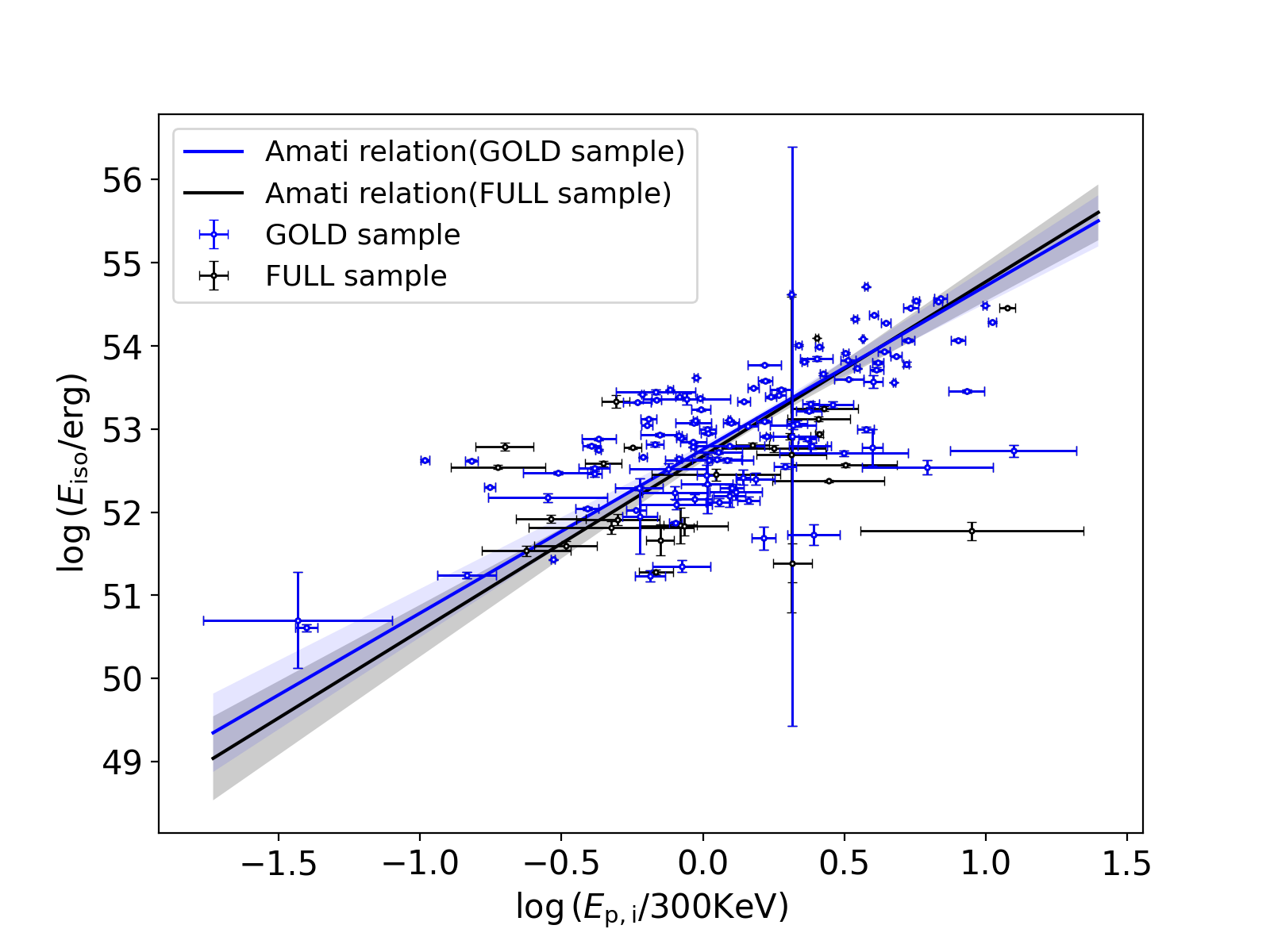}
	\caption{The best-fit calibration of Amati relation by BCES-orthogonal method with the FULL sample (\emph{black cycles}) and the GOLD sample (\emph{blue cycles}) at all $z$ by assuming the $\Lambda$CDM model from Pantheon+ ($\Omega_m = 0.334, H_0 = 73.6 \rm km/s/Mpc$)\citep{Brout2022}. FULL/BCES-Ort: $a=52.67\pm{0.08}, b=2.09\pm{0.35}$, GOLD/BCES-Ort: $a=52.75\pm{0.09}, b=1.96\pm{0.37}$. The Spearman$'$s rank coefficient of the correlation are $\rho$ = 0.60 and $\rho$ = 0.63, and the P-value are $4.75\times10^{-17}$, $4.31\times10^{-16}$ for the FULL and GOLD samples, respectively.}
	\label{fig:bces_all_lcdm}
\end{figure}

From Table \ref{tab:amati} and Figure \ref{fig:bces}, we find the  Amati relation of GRBs at $z < 1.4$  looks significantly more dispersed and flat to most previous works. This flatness can be quantified by the surprising value of the slope obtained. In order to investigate whether this flatness and high dispersion are due to the rather small sample of GRBs at $z < 1.4$ or is a characteristic of the whole \emph{Fermi}/GBM sample, we use the fitting results with the whole FULL and GOLD samples up to high redshifts by the BCES-orthogonal method. For the uncertainties of GP reconstruction extrapolating to $z > 1.4$ are very large, we assume the $\Lambda$CDM model to calculate the best-fit calibration of Amati relation by BCES-orthogonal method with the FULL and the GOLD samples at all $z$, which are shown in Figure \ref{fig:bces_all_lcdm}.
We find the value of the slope $b$ is more close to the theoretical value of the slope with the Amati relation: 2.
The Spearman$'$s rank correlation coefficients are larger than the ones at low redshift. These results indicate that this flatness and high dispersion at low redshift for the Fermi sample are due to the rather small sample of GRBs.

In order to take into account the measurement errors in both variables, we  use the calibration results of the BCES-orthogonal  method 
to constrain cosmological models.

\section{CONSTRAINING COSMOLOGICAL MODELS}\label{sec:conDEModel}
\begin{figure}
	\centering
	\includegraphics[width=\columnwidth]{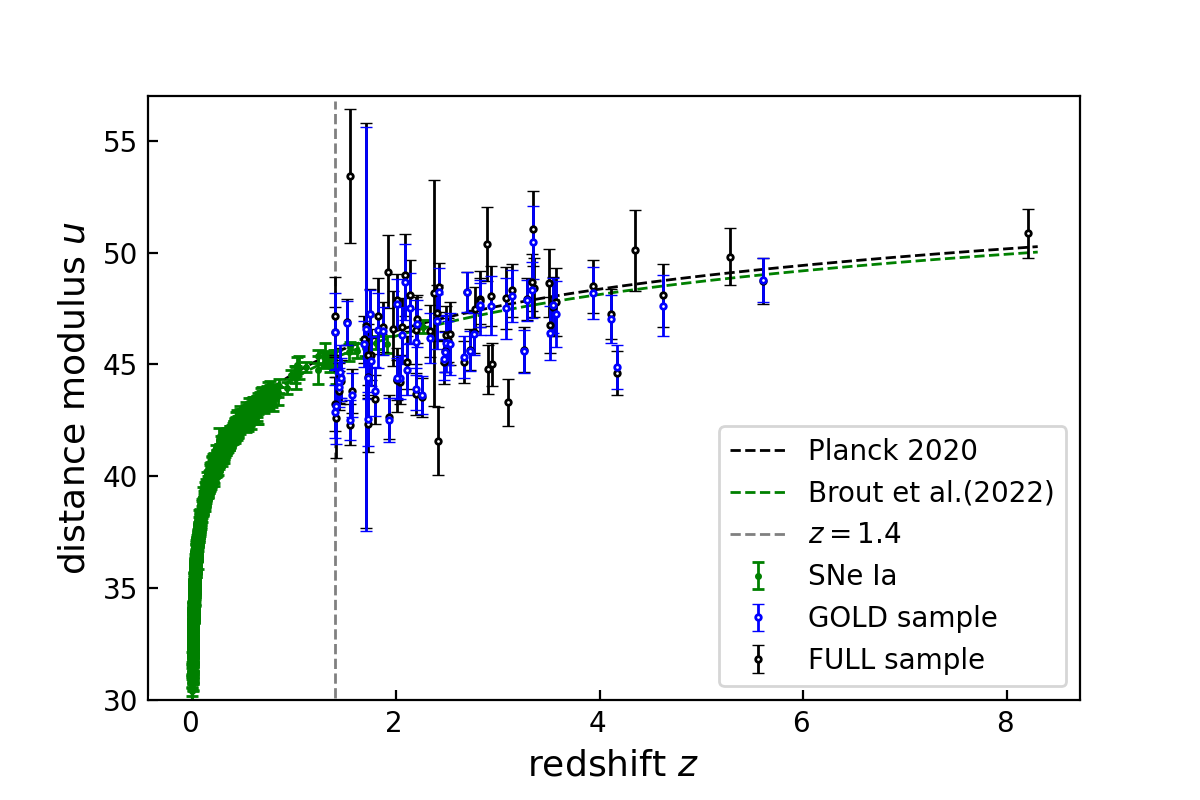}
	\caption{GRB Hubble diagram of 
		the FULL sample (\emph{black cycles}) and the GOLD sample (\emph{blue cycles}).
		The green dots represent Pantheon+ SNe Ia.
		The green dashed curve and the black dashed curve are the predicted values of distance modulus  for a flat $\Lambda$CDM model from Pantheon+ ($\Omega_m = 0.334, H_0 = 73.6 \rm km/s/Mpc$)\citep{Brout2022},  and CMB ($\Omega_m = 0.315, H_0 = 67.4 \rm km/s/Mpc$)\citep{Planck_2020}, respectively. The gray dotted line denotes $z = 1.4$.}\label{fig:HD}
\end{figure}

We assume that the calibration results of the Amati relation at low-redshift are valid at high-redshift 
to build the GRB Hubble diagram.
The Hubble diagram of the \emph{Fermi} sample is plotted in Fig. \ref{fig:HD}.
The observable distance moduli ($\mu = 5\log\frac{d_{\rm L}}{\rm Mpc} + 25$)
of GRBs can be calculated by the calibration results of the Amati relation,
\begin{equation}\label{equ:obsmu}
	\mu_{\rm GRB,obs} = 25 + \frac{5}{2}\left [a+b\log E_{\rm p,i} - \log \frac{4\pi S_{\rm bolo}}{(1+z)}\right ]
\end{equation}
The propagated uncertainties of the distance moduli is,  
\begin{equation}\scriptsize
	\begin{aligned}
		&\sigma_{\mu_{\mathrm{GRB}}}^{2}=\sigma_{\rm{cal}}^{2}+\left(\frac{\partial\mu_{\mathrm{GRB}}}{\partial E_{\mathrm{p,i}}}\right)^{2}\sigma_{E_{\mathrm{p,i}}}^{2}+\left(\frac{\partial\mu_{\mathrm{GRB}}}{\partial S_{\mathrm{bolo}}}\right)^2\sigma_{S_{\mathrm{bolo}}}^2
	\end{aligned}
\end{equation}
where  $\sigma_{\rm{cal}}^{2}$ 
is the uncertainties from calibration parameters ($a, b$) of the Amati relation, 
\begin{equation}\scriptsize
	\begin{aligned}	
		\sigma_{\rm{cal}}^{2} = \left(\frac{\partial\mu_{\mathrm{GRB}}}{\partial a}\right)^{2}\sigma_{a}^{2}+\left(\frac{\partial\mu_{\mathrm{GRB}}}{\partial b}\right)^{2}\sigma_{b}^{2} 
		+2\left(\frac{\partial\mu_{\mathrm{GRB}}}{\partial a}\right)\left(\frac{\partial\mu_{\mathrm{GRB}}}{\partial b}\right)\sigma_{ab}
	\end{aligned}
\end{equation}
here  $\sigma_{ab}$ 
is the non-diagonal matrix elements of $\mathbf{cov}$.

We use the GRB data in the Hubble diagram at high-redshift to constrain cosmological models.
The $\chi^{2}$ of the GRB data is
\begin{equation}\label{equ:chigrb}
	\chi_{\mathrm{GRB}}^2=\Delta\mu_{\mathrm{GRB}}\mathbf{C_{\rm GRB}^{-1}}\Delta\mu_{\mathrm{GRB}}
\end{equation}
here $\mathbf{C_{\rm GRB}}$ is the diagonal matrix containing of $\sigma_{\mu}^2$ \citep{Montiel2021} for simplicity, 
$\Delta\mu_{\mathrm{GRB}} =\mu_{\rm GRB,obs}-\mu_{\rm th}$ is the vector of the differences between the observable and theoretical value of the distance moduli for the GRBs.
The theoretical distance modulus is $\mu_{\rm th}(z)\equiv 5\log_{10}D_L(z)+\mu_0$,
where $\mu_{0} \equiv 42.38 - 5\log{h}$, $h$ is the Hubble constant $H_{\mathrm{0}}$ in units $\mathrm{100km/s/Mpc}$, and the unanchored luminosity
distance is
\begin{equation}\label{equ:DL}
	D_L(z)=(1+z)\int_0^z\frac{dz'}{E(z')}
\end{equation}
here $E(z)=\frac{H(z)}{H_0}$ is the unanchored Hubble parameter which can be determined  by the dark energy (DE) models.
Assuming a flat universe, we consider $\Lambda$CDM model, $w$CDM model, and the well-known  Chevallier-Polarski-Linder (CPL) model \citep{CP2001,L2003} in this work.
The Friedmann equation is given by
\begin{equation}\scriptsize
	\label{equ:model}
	E(z)\equiv\begin{cases}
		\sqrt{\Omega_m(1+z)^3+\Omega_{\mathrm{DE}}}, \rm\Lambda CDM~model \\
		\\
		\sqrt{\Omega_m(1+z)^3+\Omega_{\mathrm{DE}}(1+z)^{3(1+w_0)}}, w\rm CDM~model \\
		\\
		\sqrt{\Omega_m(1+z)^3+\Omega_{\text{DE}}(1+z)^{3(1+w_0+w_a)}\exp\left(-\frac{3w_az}{1+z}\right)}, \rm CPL\\
	\end{cases}
\end{equation}

We also combine the Pantheon+ SNe Ia sample \citep{Scolnic2022} to constrain cosmological models with the GRB sample.
The $\chi^{2}$ of the SNe Ia is 
\begin{equation}\label{equ:chisne}
	\chi_{\mathrm{SN}}^2=\Delta\mu_{\mathrm{SN}}^{\mathrm{T}}\mathbf{C}^{-1}_{\rm SN}\Delta\mu_{\mathrm{SN}}
\end{equation}
where $\Delta\mu_{\mathrm{SN}}\equiv\mu_{\mathrm{th}}-\mu_{\rm SN,obs}$ is the module of the vector of residuals, $\rm\textbf{C}_{\rm SN}$ is the covariance matrix \citep{Brout2022} of 1701 SNe Ia.
The total $\chi^{2}$ with SNe Ia and GRB is given by
\begin{equation}\label{equ:total}
	\chi^{2}_{\rm tot}  = \chi_{\mathrm{GRB}}^{2} + \chi_{\mathrm{SNe}}^{2}
\end{equation}

We use the MCMC method of Python package emcee \citep{Foreman-Mackey_2013} to minimize the $\chi^{2}$.
In order to constrain DE models, we use the 15 years of the \emph{Fermi}-GBM GRBs with the FULL sample and the GOLD sample 
at high-redshift $z\ge1.4$.
For comparison, we also use GRB data at high-redshift from M2021 \citep{Montiel2021} with the 74 GRBs (2008-2019)  at $0.117\le z\le5.283$ 
in the same calibration method through Gaussian process from the latest OHD.

\begin{figure}
	\centering
	\includegraphics[width=0.5\columnwidth]{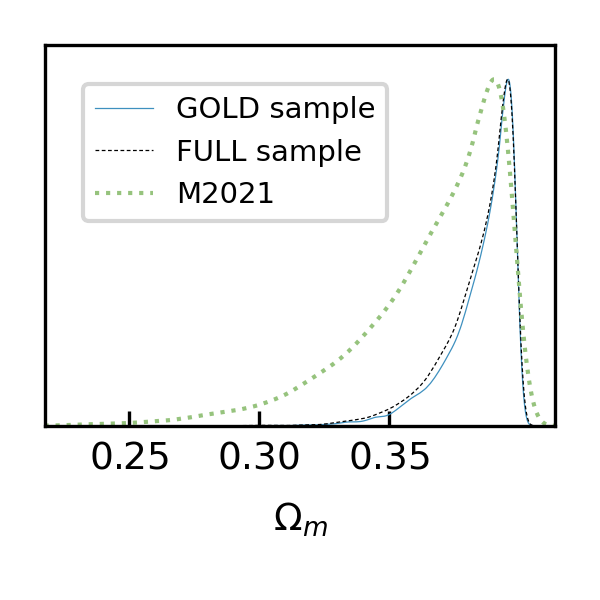}
	\caption{Constraints on parameters of $\Omega_{\rm m}$ for the flat $\Lambda$CDM model with GRBs at $z\ge1.4$ from the FULL sample (\emph{black dashed curves}), the GOLD sample (\emph{blue curves}), and the M2021 sample (\emph{green dotted curves}), respectively.}\label{fig:lCDM}
\end{figure}

\begin{figure}
	\centering
	\includegraphics[width=0.7\columnwidth]{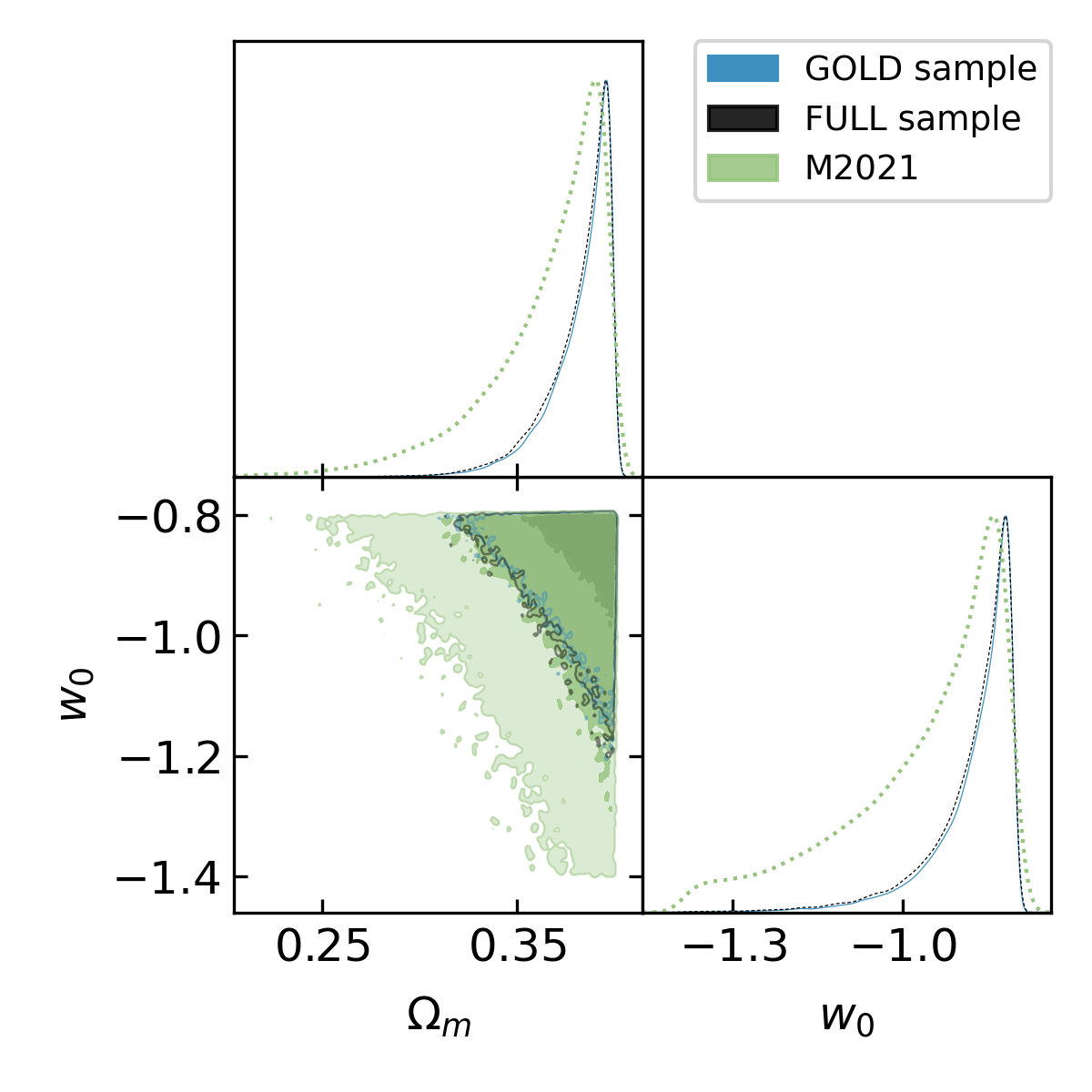}
	\caption{Constraints on parameters of $\Omega_{\rm m}$ and $w_{0}$ for the flat $w$CDM model with GRBs at $z\ge1.4$ from the FULL sample (\emph{black dashed curves}), the GOLD sample (\emph{blue curves}), and the M2021 sample (\emph{green dotted curves}), respectively.}\label{fig:wCDM}
\end{figure}

\begin{figure}
	\centering
	\includegraphics[width=0.8\columnwidth]{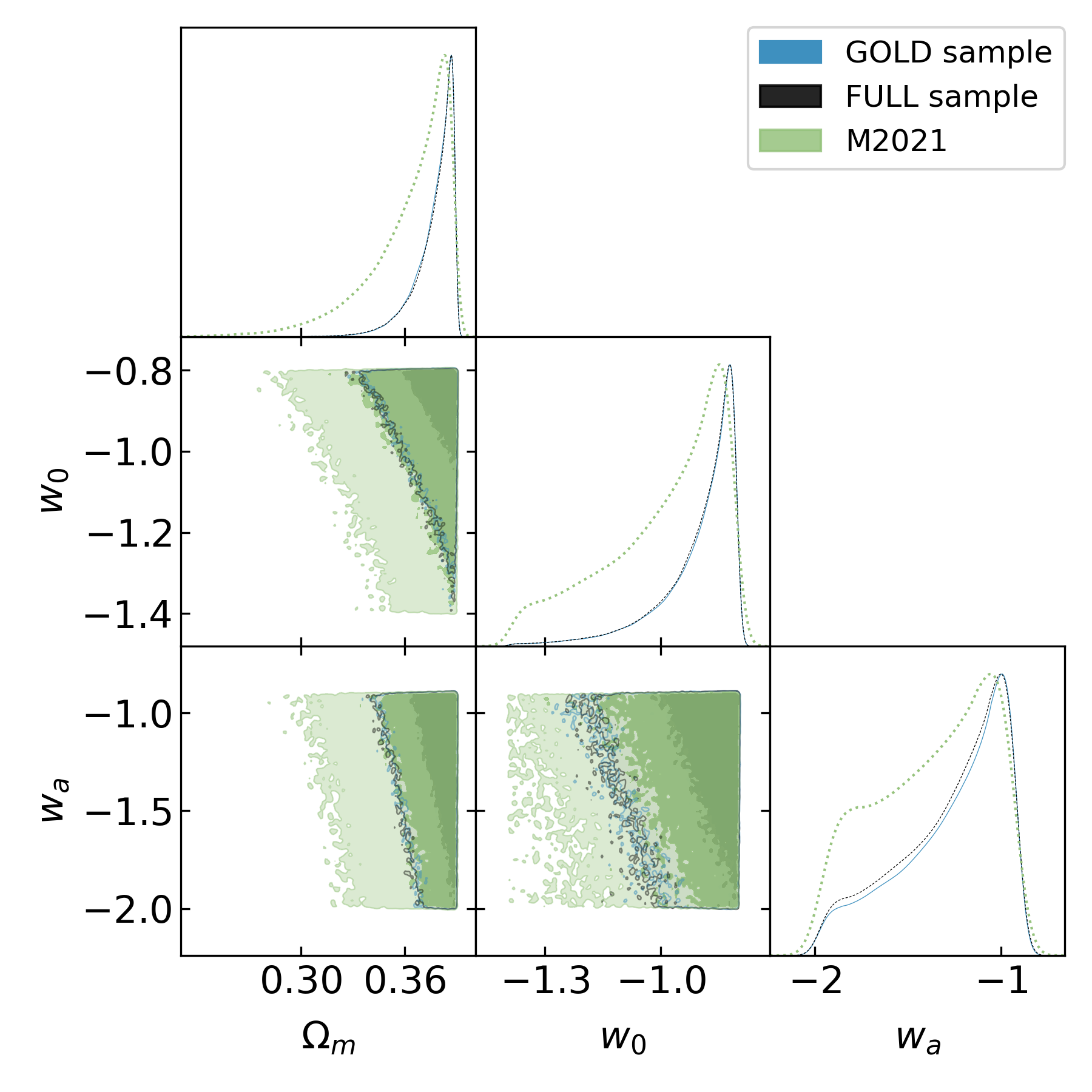}
	\caption{Constraints on parameters of $\Omega_{\rm m}$, $w_{0}$ and $w_{a}$ for the flat CPL model with GRBs at $z\ge1.4$ from the FULL sample (\emph{black dashed curves}), the GOLD sample (\emph{blue curves}), and the M2021 sample (\emph{green dotted curves}) respectively.}\label{fig:CPL}
\end{figure}

We set $H_0=73\rm km/s/Mpc$\textbf{\footnote{If just to make the comparison with the cosmological parameters obtained by joining GRBs with SNe more straightforward, it would be better to assume $H_0=73\rm km/s/Mpc$ for the derivation of cosmological parameters based on GRBs only. Of course, this should not impact significantly the result, but it would allow a more consistent comparison.}} to constrain only with GRBs at $z\ge1.4$ including the GOLD, FULL and M2021 sample.
The results are summarized in Table \ref{tab:cosm}.
For the flat $\Lambda$CDM model (Figure \ref{fig:lCDM}), we obtain
$\Omega_{\rm m} = 0.386^{+0.014}_{-0.004}$ and $\Omega_{\rm m} = 0.385^{+0.015}_{-0.005}$with the GOLD and FULL sample, respectively;
for the flat $w$CDM model (Figure \ref{fig:wCDM}), we obtain 
$w_0 = -0.87^{+0.07}_{-0.02}$ and $w_0 = -0.88^{+0.08}_{-0.02}$ with the GOLD and FULL sample, respectively;
for the CPL model (Figure \ref{fig:CPL}), we obtain 
$w_a = -1.25^{+0.35}_{-0.13}$ and $w_a = -1.24^{+0.33}_{-0.13}$ with the GOLD and FULL sample,
which favours a possible DE evolution at the $1\sigma$ confidence region for all cases only with GRBs at $z\ge1.4$.
Our results with the GOLD and FULL sample are almost identical, which are more stringent than the results of M2021 sample.
We find that the values of $\Omega_{\rm m}$ with the GOLD and FULL sample for the flat $\Lambda$CDM model is larger than the result of M2021 sample: $\Omega_{\rm m}=0.370^{+0.030}_{-0.009}$, which are much less than ones obtained in previous analyses only with A219 and A118 GRBs at high-redshift: 
$\Omega_{\rm m}\sim0.5$ \citep{Liang2022}.

\begin{figure}
	\centering
	\includegraphics[width=0.7\columnwidth]{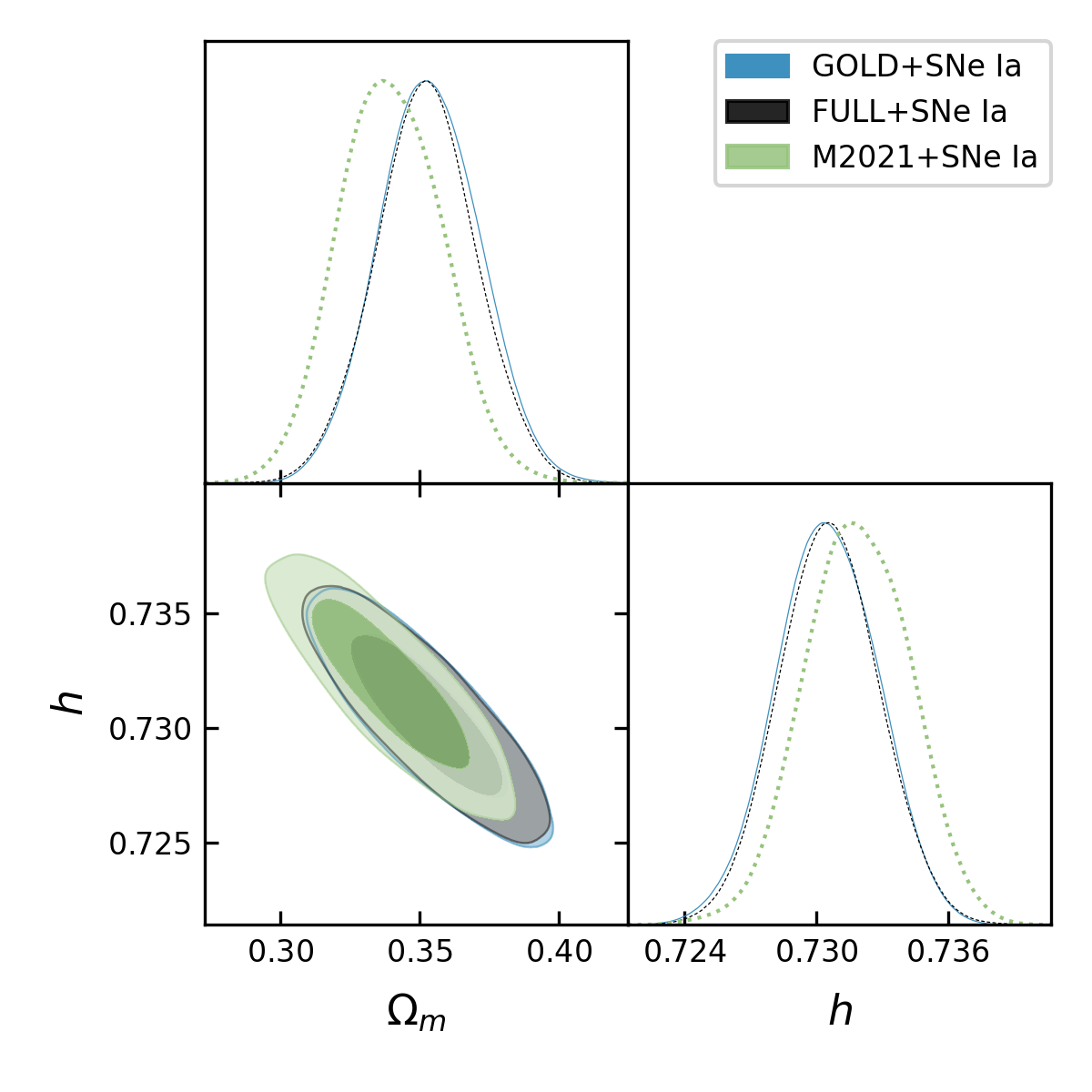}
	\caption{Constraints on parameters of $\Omega_{\rm m}$, $h$ for the flat $\Lambda$CDM model with SNe Ia and GRBs at $z\ge1.4$ from the FULL sample (\emph{black dashed curves}), the GOLD sample (\emph{blue curves}), and the M2021 sample (\emph{green dotted curves}), respectively.}\label{fig:lCDMgs}
\end{figure}
\begin{figure}
	\centering
	\includegraphics[width=0.8\columnwidth]{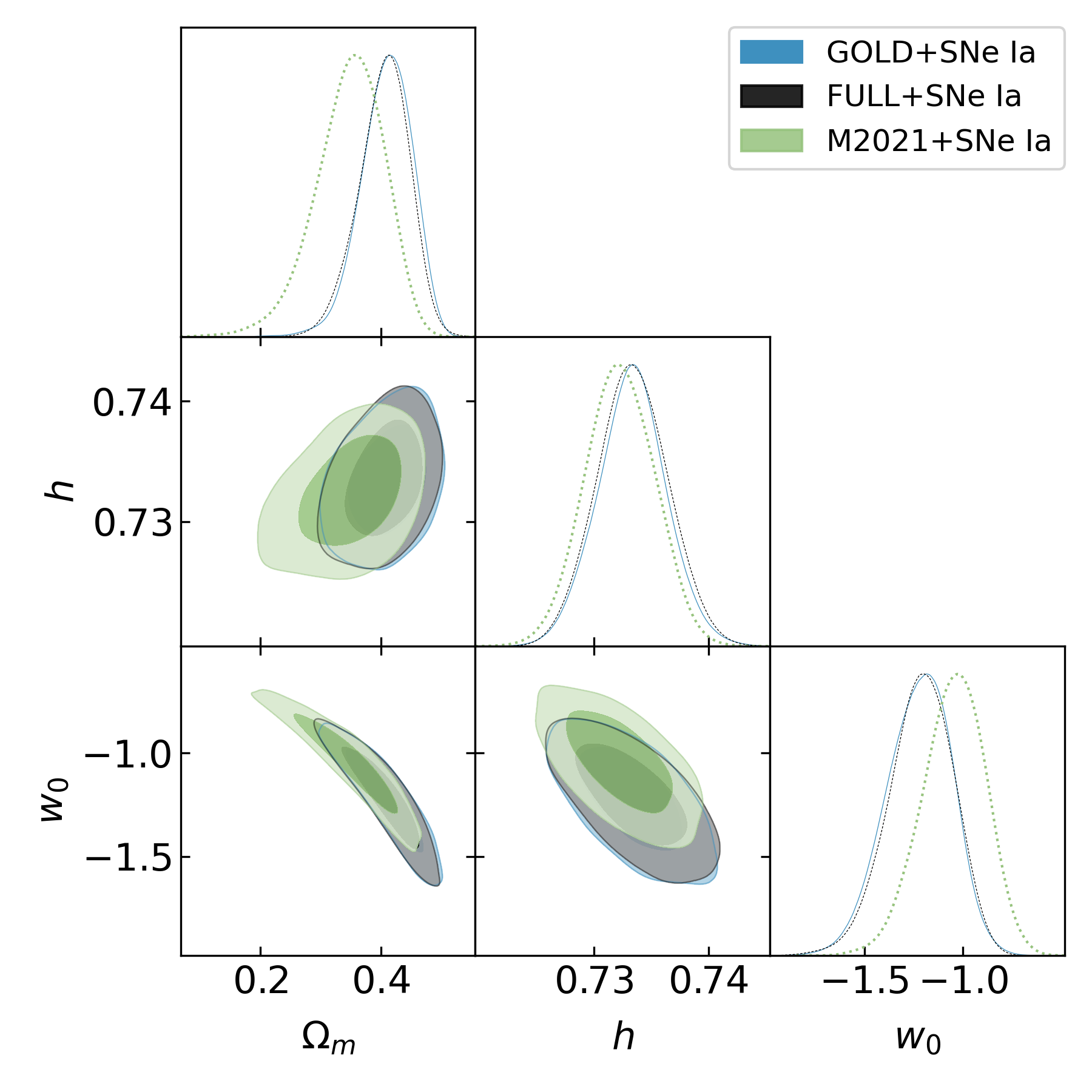}
	\caption{Constraints on parameters of $\Omega_{\rm m}$, $h$, and $w_{0}$ for the flat $w$CDM model with SNe Ia and GRBs at $z\ge1.4$ from the FULL sample (\emph{black dashed curves}), the GOLD sample (\emph{blue curves}), and the M2021 sample (\emph{green dotted curves}) respectively.}\label{fig:wCDMgs}
\end{figure}
\begin{figure}
	\centering
	\includegraphics[width=0.9\columnwidth]{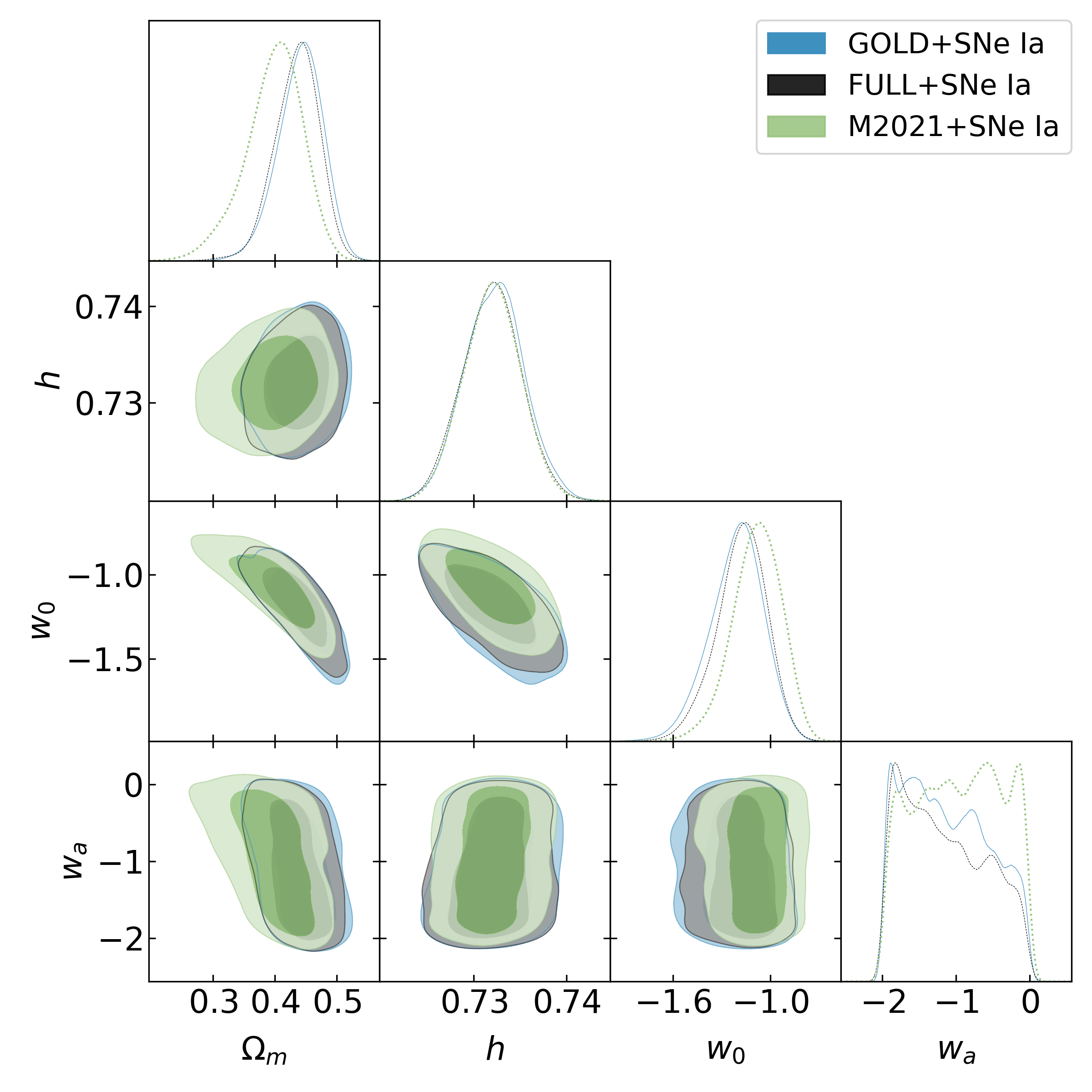}
	\caption{Constraints on parameters of $\Omega_{\rm m}$, $h$, $w_{0}$ and $w_{\rm a}$ for the flat CPL model with SNe Ia and GRBs at $z\ge1.4$ from the FULL sample (\emph{black dashed curves}), the GOLD sample (\emph{blue curves}), and the M2021 sample (\emph{green dotted curves}) respectively.}\label{fig:CPLgs}
\end{figure}

The joint constraints with Pantheon+ SNe Ia and the GOLD, FULL and M2021 sample at high-redshift are summarized in Table \ref{tab:cosmGRBSN}.
For the flat $\Lambda$CDM model (Figure \ref{fig:lCDMgs}):
$\Omega_{\rm m} = 0.354\pm0.018, H_0 = 73.05\pm0.2$ and
$\Omega_{\rm m} = 0.353\pm0.018, H_0 = 73.06\pm0.2$
with the GOLD and FULL sample, respectively, which are slightly different with ones obtained with the Pantheon+ sample $\Omega_{\rm m} = 0.334\pm0.018, H_0 = 73.6\pm1.1$ for the flat $\Lambda$CDM model \citep{Brout2022};
for the flat $w$CDM model (Figure \ref{fig:wCDMgs}):
$w_0 = -1.22^{+0.18}_{-0.15}$ and $w_0 = -1.21^{+0.18}_{-0.15}$
with the GOLD and FULL sample, respectively;
for the CPL model (Figure \ref{fig:CPLgs}):
$w_{a} = -1.12^{+0.45}_{-0.83}$ and $w_{a} = -1.16^{+0.66}_{-0.80}$ with the GOLD and FULL sample, respectively.
Our results with the GOLD and FULL sample are almost identical, which are more stringent than the results of M2021 sample.
We find our results with the GOLD, FULL sample and Pantheon+ sample that are quite different with the previous analyses obtained from A118 sample at high-redshift and Pantheon sample: $\Omega_{\rm m} = 0.286\pm0.012, H_0 = 69.70\pm0.22$ for the flat $\Lambda$CDM model \citep{Li_2023}.

Finally, we compare  these dark energy models with Akaike Information Criterion (AIC), Bayesian Information Criterion (BIC), 
which defined by
\begin{equation}\label{equ:AIC}
	\text{AIC}=-2\ln\mathcal{L}_{\rm max}+2k,~
	\text{BIC}=-2\ln\mathcal{L}_{\rm max}+k\ln N
	\
\end{equation}
where $\mathcal{L}_{\rm max}$ is the maximum likelihood.
The value of $\Delta\text{AIC}$ and $\Delta\text{BIC}$ are given by
\begin{equation}\label{equ:dIC}
	\Delta\text{AIC}=\Delta\chi_{\rm min}^2+2\Delta k,~
	\Delta\text{BIC}=\Delta\chi_{\rm min}^2+\Delta k\ln N
\end{equation}
In the Gaussian cases, $\begin{aligned}\chi_{\rm min}^2=-2\ln\mathcal{L}_{\rm max}\end{aligned}$.
The results of the value of $\Delta\text{AIC}$ and $\Delta\text{BIC}$ 
with respect to the reference model (the $\rm \Lambda$CDM model) are summarized in Table \ref{tab:cosm} and \ref{tab:cosmGRBSN}.
We find that $\rm \Lambda$CDM model is preferred with respect to the $w$CDM and the CPL models, which is consistent with the previous analyses \citep{Amati_2019, Montiel2021} and \citep{Li_2023}.

\begin{table*}
	\caption{The constraints only with GRBs at $z > 1.4$ in the FULL sample, the GOLD sample, and the M2021 sample for the flat $\Lambda$CDM model, the flat $w$CDM model, and the CPL model  by setting $H_0=73\rm km/s/Mpc$.}
	\label{tab:cosm}
	\centering
	\begin{tabular}{llcccccc}
		\hline\hline
		Data Set & Models & \boldmath{$\Omega_{\rm m}$} & \boldmath{$w_0$} & $w_{a}$ & $\Delta$AIC & $\Delta$BIC \\ \hline
		\multirow{3}{*}{GOLD sample} & $\Lambda$CDM & $0.386^{+0.014}_{-0.004}$ & - & - & 0 & 0 \\ \cline{2-7}
		& $w$CDM       & $0.383^{+0.017}_{-0.005}$ & $-0.87^{+0.07}_{-0.02}$ & - & 0.55 & 1.56  \\ \cline{2-7}
		& CPL          & $0.379^{+0.011}_{-0.002}$ & $-0.89^{+0.09}_{-0.02}$ & $-1.25^{+0.35}_{-0.13}$ & 1.15 & 2.91  \\ \hline
		
		\multirow{3}{*}{FULL sample} & $\Lambda$CDM & $0.385^{+0.015}_{-0.005}$ & - & - & 0 & 0  \\ \cline{2-7}
		& $w$CDM       & $0.382^{+0.018}_{-0.005}$ & $-0.88^{+0.08}_{-0.02}$ & - & 0.77 & 2.90  \\ \cline{2-7}
		& CPL          & $0.378^{+0.012}_{-0.003}$ & $-0.90^{+0.10}_{-0.02}$ & $-1.24^{+0.33}_{-0.13}$ & 1.58 & 3.86  \\ \hline
		
		\multirow{3}{*}{M2021} & $\Lambda$CDM & $0.368^{+0.032}_{-0.011}$ & - & - & 0 & 0  \\ \cline{2-7}
		& $w$CDM       & $0.365^{+0.035}_{-0.010}$ & $-0.96^{+0.16}_{-0.05}$ & - & 1.15 & 3.25  \\ \cline{2-7}
		& CPL          & $0.363^{+0.027}_{-0.007}$ & $-0.99^{+0.19}_{-0.07}$ & $-1.36^{+0.43}_{-0.22}$ & 2.41 & 4.48  \\ \hline
		
	\end{tabular}
\end{table*}

\begin{table*}
	\caption{The constraints on parameters of $\Omega_{\rm m}$, $h$, $w_{0}$ and $w_{a}$ for the flat $\Lambda$CDM model, the flat $w$CDM model, and the CPL model with SNe Ia and GRBs at $z > 1.4$ in  the FULL sample, the GOLD sample, and the M2021 sample.}
	\label{tab:cosmGRBSN}
	\centering
	\begin{tabular}{llcccccc}
		\hline\hline
		Data Set & Models & $\Omega_{\rm m}$ & $h$ & $w_0$ & $w_{a}$ & $\Delta$AIC & $\Delta$BIC \\ \hline
		\multirow{3}{*}{GOLD+SNe Ia} & $\Lambda$CDM & $0.354^{+0.018}_{-0.018}$ & $0.7305^{+0.0023}_{-0.0023}$ & - & - & 0 & 0\\ \cline{2-8}
		& $w$CDM & $0.410^{+0.047}_{-0.037}$ & $0.7333^{+0.0030}_{-0.0030}$ & $-1.22^{+0.18}_{-0.15}$ & - & 0.28 & 5.69\\ \cline{2-8}
		& CPL & $0.441^{+0.040}_{-0.032}$ & $0.7322^{+0.0032}_{-0.0032}$ & $-1.21^{+0.18}_{-0.15}$ & $-1.12^{+0.45}_{-0.83}$ & 0.44 & 11.25 \\ \hline
		
		\multirow{3}{*}{FULL+SNe Ia} & $\Lambda$CDM & $0.353^{+0.018}_{-0.018}$ & $0.7306^{+0.0023}_{-0.0023}$ & - & - & 0 & 0\\ \cline{2-8}
		& $w$CDM & $0.406^{+0.046}_{-0.036}$ & $0.7334^{+0.0031}_{-0.0031}$ & $-1.21^{+0.18}_{-0.15}$ & - & 0.45 & 5.87\\ \cline{2-8}
		& CPL & $0.435^{+0.039}_{-0.032}$ & $0.7319^{+0.0032}_{-0.0032}$ & $-1.18^{+0.17}_{-0.13}$ & $-1.16^{+0.66}_{-0.80}$ & 0.69 & 11.53 \\ \hline
		
		\multirow{3}{*}{M2021+SNe Ia} & $\Lambda$CDM & $0.340^{+0.018}_{-0.018}$ & $0.7319^{+0.0024}_{-0.0024}$ & - & - & 0 & 0\\ \cline{2-8}
		& $w$CDM & $0.348^{+0.064}_{-0.048}$ & $0.7323^{+0.0030}_{-0.0030}$ & $-1.05^{+0.17}_{-0.14}$ & - & 1.95 & 7.35\\ \cline{2-8}
		& CPL & $0.398^{+0.052}_{-0.037}$ & $0.7320^{+0.0031}_{-0.0031}$ & $-1.08^{+0.17}_{-0.13}$ & $-0.96^{+0.58}_{-0.58}$ & 3.55 & 14.34 \\ \hline
		
	\end{tabular}
\end{table*}



\section{CONCLUSION AND DISCUSSIONS}\label{sec:conclusion}
In the present work, we compile a \emph{Fermi} sample of the \emph{long} GRBs from  \emph{Fermi} observations with 15 years of the Fermi-GBM catalogue with identified redshift, in which the GOLD sample contains 123 long GRBs at $z\le5.6$ and the FULL sample contains 151 long GRBs with redshifts at $z\le8.2$.
The Amati relation are calibrated from the latest OHD  with the CC method  using a Gaussian Process at $z<1.4$ to obtain GRBs  at $z\ge1.4$  which can be used to constrain cosmological models.
By the MCMC method  with the GRBs in the GOLD and FULL sample at $z\ge1.4$  and the Pantheon+ sample, 
we obtain $\Omega_{\rm m} = 0.354\pm0.018, H_0 = 73.05\pm0.2$ and $\Omega_{\rm m} = 0.353\pm0.018, H_0 = 73.06\pm0.2$  for the flat $\Lambda$CDM model;
$w_0 = -1.22^{+0.18}_{-0.15}$ and $w_0 = -1.21^{+0.18}_{-0.15}$ for the flat $w$CDM model; $w_{a} = -1.12^{+0.45}_{-0.83}$ and $w_{a} = -1.16^{+0.66}_{-0.80}$  for the CPL model.
We also find that $\rm \Lambda$CDM model is preferred with respect to the $w$CDM and the CPL models.

It should  be note that there is a debate as to whether the Amati relation is an intrinsic effect or the result of detection biases \citep{Butler2007,Butler2009,Butler2010} and the instrumental selection biases \citep{Ghirlanda2008,Nava2012}, which cannot be responsible for the existence of the spectral-energy correlations.
Our sample in this work originate from a single catalogue, thus avoiding selection biases and other instrument-associated systematics. Moreover, the Amati relation could arguably evolve with redshift, which have been discussed in many works \citep{Demianski_2017a,Tang2021,Khadka2021}.
\cite{Liu2022a,Liu2022b} proposed the improved Amati relation by accounting for evolutionary effects via copula to found that a redshift evolutionary correlation is slightly favored.
\cite{Jia_2022} found no statistically significant evidence for the redshift evolution with the Amati relation. 
However, \cite{Frontera2012} found that the time-resolved relation (the Yonetoku relation: $E_{\rm p,i}$-$L_{\rm p, iso}$ correlation) is a clear demonstration that the origin of the correlation is physical, and the Amati relation and the Yonetoku relation likely not the result of selection effects. Moreover, \cite{Dainotti_2018} have given a review of the evidences showing that the Amati relation is NOT affected by relevant selection effects and biases.
Further examinations of possible evolutionary effects should be required for considering GRBs as standard candles for a cosmological probe.

For the GRB satellites, the Konus-Wind (KW) experiment plays an important role in GRB studies thanks to its unique set of characteristics: the spacecraft orbit in interplanetary space that provides an exceptionally stable background; the continuous coverage of the full sky by two omnidirectional detectors; and the broad energy range (20 keV-15MeV, triggered mode) since 1994.
The \emph{Swift} satellite was launched in 2004, with the Burst Alert Telescope (BAT) instrument which is capable of detecting energies up to 150 keV \citep{Gehrels2004}.
In the future, we are looking forward to the observations by the French-Chinese satellite spacebased multi-band astronomical variable objects monitor (SVOM) \citep{Wei_2016} and the Transient High-Energy Sky and Early Universe Surveyor (THESEUS) \citep{Amati_2018}.

\cite{Atteia2017} used the sample of GRBs with redshift detected by Fermi/GBM (52 GRBs with a cutoff at GRB160629) and KW to investigate the maximum isotropic energy of GRBs;
\cite{Tsvetkova2021} presented a systematic study of GRBs with reliable redshift estimates detected simultaneously by the KW and the Swift/BAT during 2005-2018.
\cite{Oates2023} provided an overview of the long GRBs observed by the Ultra-Violet/Optical Telescope (UVOT) onboard \emph{Swift}, and review the major discoveries that have been achieved by the \emph{Swift}/UVOT over the last 18 years during 2004-2022.
Along with the GRB sample from KW, Swift, and Fermi, as well as SVOM and THESEUS, GRBs could be used as an additional choice to help us explore universe using GRBs at high-redshift.

\section*{ACKNOWLEDGMENTS}
We are grateful to Dr. Ariadna Montiel, Prof. $\emptyset$yvind Gr$\emptyset$n for useful discussions.
We also thank the referee for helpful comments and constructive suggestions.
This project was supported by the Guizhou Provincial Science and Technology Foundations (QKHJC-ZK[2021] Key 020 and QKHJC-ZK[2024] general 443).
\section*{DATA AVAILABILITY}
Data are available at the following references:
the Fermi GRB sample are from this work and references therein,
the Pantheon+ SNe Ia sample from \cite{Scolnic2022},
and the latest OHD obtained with the CC method from \cite{Moresco_2020,Moresco_2022} and \cite{LZL2023}.



\bsp	
\label{lastpage}

\appendix
\section{Data Sample Table} 
\onecolumn
\begin{ThreePartTable}\scriptsize
	\begin{TableNotes}
		\item[$a$] The spectral parameters of 18 GRBs are taken from the GCN.
		\item[$b$] The spectral parameters of 9 GRBs are obtained in this work.
	\end{TableNotes}
	\begin{longtable}{lllllllllcc}
		\caption{The GOLD sample (\aznum LGRBs) table.}\label{tab:LGRBs} \\
		\hline\hline
		GRB & $z$ & $T_{90}$ & $\alpha$ & $\sigma_{\alpha}$ & $\beta$ & $\sigma_{\beta}$ & $E_{\rm p,i}(\rm keV)$ & $\sigma_{E_{\rm p,i}}(\rm keV)$ & $S_{\rm bolo}(\rm erg/cm^2)$ & $\sigma_{S_{\rm bolo}}(\rm erg/cm^2)$ \\
		\hline
		\endfirsthead
		\caption*{The GOLD sample (\aznum LGRBs) table. (continued)}\\
		\hline\hline
		GRB & $z$ & $T_{90}$ & $\alpha$ & $\sigma_{\alpha}$ & $\beta$ & $\sigma_{\beta}$ & $E_{\rm p,i}(\rm keV)$ & $\sigma_{E_{\rm p,i}}(\rm keV)$ & $S_{\rm bolo}(\rm erg/cm^2)$ & $\sigma_{S_{\rm bolo}}(\rm erg/cm^2)$ \\
		\hline
		\endhead
		\hline
		\endfoot
		\insertTableNotes
		\endlastfoot
		080804  & 2.2045 & 24.7  & -5.175E-01 & 7.985E-02 & -1.905E+00 & 8.533E-02 & 7.088E+02 & 7.356E+01 & 1.357E-05 & 6.537E-07 \\
		080810  & 3.35   & 75.2  & -1.090E+00 & 4.453E-02 & -5.982E+00 & 9.814E+01 & 2.564E+03 & 3.734E+02 & 1.165E-05 & 4.552E-07 \\
		080916A & 0.689  & 46.34 & -7.807E-01 & 1.063E-01 & -1.774E+00 & 4.619E-02 & 1.786E+02 & 3.454E+01 & 1.473E-05 & 7.589E-07 \\
		081121  & 2.512  & 41.99 & -4.351E-01 & 1.148E-01 & -2.096E+00 & 9.481E-02 & 5.652E+02 & 5.074E+01 & 1.985E-05 & 8.038E-07 \\
		081221  & 2.26   & 29.7  & -8.387E-01 & 2.245E-02 & -3.675E+00 & 4.704E-01 & 2.833E+02 & 4.328E+00 & 3.286E-05 & 2.109E-07 \\
		081222  & 2.77   & 18.88 & -8.444E-01 & 4.478E-02 & -2.300E+00 & 1.188E-01 & 5.550E+02 & 3.179E+01 & 1.434E-05 & 4.283E-07 \\
		090113  & 1.7493 & 17.41 & -1.183E+00 & 1.420E-01 & -2.065E+00 & 5.152E-01 & 3.922E+02 & 8.461E+01 & 2.194E-06 & 4.937E-07 \\
		090323  & 3.57   & 133.9 & -1.183E+00 & 1.145E-02 & -2.354E+00 & 1.470E-01 & 2.073E+03 & 1.078E+02 & 1.364E-04 & 2.690E-06 \\
		090328A & 0.736  & 61.7  & -1.083E+00 & 1.657E-02 & -2.390E+00 & 2.323E-01 & 1.129E+03 & 7.598E+01 & 6.532E-05 & 4.546E-06 \\
		090424  & 0.544  & 14.14 & -1.022E+00 & 1.625E-02 & -2.763E+00 & 1.704E-01 & 2.470E+02 & 6.140E+00 & 5.427E-05 & 1.593E-06 \\
		090516A & 4.109  & 123.1 & -1.456E+00 & 4.396E-02 & -2.460E+00 & 8.406E-01 & 7.563E+02 & 1.012E+02 & 2.068E-05 & 1.338E-06 \\
		090618  & 0.54   & 112.4 & -1.114E+00 & 1.308E-02 & -2.239E+00 & 2.007E-02 & 2.295E+02 & 5.061E+00 & 3.698E-04 & 3.627E-06 \\
		090902B & 1.822  & 19.33 & -1.008E+00 & 4.173E-03 & -9.722E+00 & 4.105E+02 & 2.978E+03 & 4.590E+01 & 3.547E-04 & 2.618E-06 \\
		090926A & 2.1062 & 13.76 & -8.480E-01 & 8.619E-03 & -2.378E+00 & 4.558E-02 & 1.037E+03 & 1.814E+01 & 1.887E-04 & 2.308E-06 \\
		090926B & 1.24   & 64    & 2.346E-01  & 1.057E-01 & -3.343E+00 & 4.482E-01 & 1.844E+02 & 5.736E+00 & 1.087E-05 & 1.754E-07 \\
		091003A & 0.8969 & 20.22 & -1.072E+00 & 2.277E-02 & -2.215E+00 & 1.480E-01 & 7.023E+02 & 5.047E+01 & 3.435E-05 & 2.238E-06 \\
		091020  & 1.71   & 24.26 & -1.245E+00 & 8.075E-02 & -2.454E+00 & 3.775E+01 & 6.188E+02 & 1.323E+02 & 1.058E-05 & 8.474E-05 \\
		091024  & 1.092  & 93.95 & -1.329E+00 & 8.007E-02 & -6.127E+00 & 2.440E+02 & 3.758E+03 & 1.932E+03 & 1.653E-05 & 2.810E-06 \\
		091127  & 0.49   & 8.701 & -1.254E+00 & 6.618E-02 & -2.216E+00 & 2.009E-02 & 5.283E+01 & 2.310E+00 & 3.034E-05 & 4.794E-07 \\
		091208B & 1.063  & 12.48 & -6.152E-01 & 2.236E-01 & -1.923E+00 & 4.911E-02 & 9.227E+01 & 2.631E+01 & 9.474E-06 & 4.674E-07 \\
		100414A & 1.368  & 26.5  & -6.242E-01 & 1.384E-02 & -3.534E+00 & 1.245E+00 & 1.571E+03 & 3.639E+01 & 1.176E-04 & 8.121E-06 \\
		100615A & 1.398  & 37.38 & -9.064E-01 & 1.560E-01 & -1.803E+00 & 3.070E-02 & 1.284E+02 & 1.800E+01 & 1.431E-05 & 3.925E-07 \\
		100728A & 1.567  & 165.4 & -5.097E-01 & 2.151E-02 & -2.542E+00 & 9.954E-02 & 6.518E+02 & 1.686E+01 & 1.564E-04 & 4.154E-06 \\
		100814A & 1.44   & 150.5 & -2.419E-01 & 1.001E-01 & -2.437E+00 & 2.555E-01 & 3.116E+02 & 2.130E+01 & 1.775E-05 & 1.387E-06 \\
		100816A & 0.8035 & 2.045 & -3.178E-01 & 7.378E-02 & -2.733E+00 & 2.726E-01 & 2.401E+02 & 1.277E+01 & 4.101E-06 & 2.296E-07 \\
		100906A & 1.727  & 110.6 & -9.263E-01 & 2.344E-01 & -1.861E+00 & 1.001E-01 & 2.043E+02 & 6.617E+01 & 3.543E-05 & 2.381E-06 \\
		101213A & 0.414  & 45.06 & -9.755E-01 & 6.427E-02 & -2.374E+00 & 7.771E-01 & 4.931E+02 & 4.786E+01 & 1.049E-05 & 3.411E-06 \\
		110213A & 1.46   & 34.31 & -1.563E+00 & 4.812E-02 & -4.870E+00 & 0.000E+00 & 2.769E+02 & 2.961E+01 & 1.207E-05 & 3.160E-07 \\
		110402A & 0.854  & 35.65 & -1.352E+00 & 7.834E-02 & -9.306E+00 & 1.427E+04 & 1.865E+03 & 9.950E+02 & 1.718E-05 & 3.393E-06 \\
		110731A & 2.83   & 7.485 & -8.686E-01 & 3.117E-02 & -2.436E+00 & 2.741E-01 & 1.234E+03 & 6.479E+01 & 2.812E-05 & 1.502E-06 \\
		111228A & 0.714  & 99.84 & -1.584E+00 & 8.630E-02 & -2.446E+00 & 8.507E-02 & 4.550E+01 & 2.343E+00 & 2.931E-05 & 9.430E-07 \\
		120119A & 1.728  & 55.3  & -9.550E-01 & 3.240E-02 & -2.366E+00 & 1.603E-01 & 4.986E+02 & 2.851E+01 & 4.823E-05 & 2.189E-06 \\
		120326A & 1.798  & 11.78 & -6.790E-01 & 2.284E-01 & -2.335E+00 & 1.347E-01 & 1.240E+02 & 1.563E+01 & 4.005E-06 & 1.666E-07 \\
		120624B & 2.1974 & 271.4 & -9.157E-01 & 1.224E-02 & -2.216E+00 & 7.805E-02 & 2.032E+03 & 7.402E+01 & 2.824E-04 & 5.881E-06 \\
		120711A & 1.405  & 44.03 & -9.844E-01 & 8.513E-03 & -2.796E+00 & 9.103E-02 & 3.169E+03 & 1.016E+02 & 3.606E-04 & 5.699E-06 \\
		120712A & 4.1745 & 22.53 & 6.105E+00  & 6.714E-01 & -1.608E+00 & 3.826E-02 & 1.758E+02 & 1.996E+01 & 6.096E-06 & 1.320E-07 \\
		120716A & 2.486  & 226   & -8.946E-01 & 5.233E-02 & -2.630E+00 & 2.198E-01 & 4.180E+02 & 2.252E+01 & 1.447E-05 & 4.231E-07 \\
		120811C & 2.671  & 14.34 & -7.028E-01 & 2.255E-01 & -2.839E+00 & 3.367E-01 & 2.034E+02 & 1.443E+01 & 3.917E-06 & 2.737E-07 \\
		120907A & 0.97   & 5.76  & -8.553E-01 & 2.148E-01 & -4.902E+00 & 5.076E+01 & 2.527E+02 & 5.961E+01 & 8.580E-07 & 1.440E-07 \\
		120909A & 3.93   & 112.1 & -8.436E-01 & 5.132E-02 & -1.934E+00 & 7.372E-02 & 9.843E+02 & 1.197E+02 & 1.260E-05 & 3.443E-07 \\
		121128A & 2.2    & 17.34 & -6.837E-01 & 1.195E-01 & -2.424E+00 & 9.205E-02 & 1.923E+02 & 1.232E+01 & 1.096E-05 & 2.588E-07 \\
		130215A & 0.597  & 143.7 & -1.059E+00 & 9.077E-02 & -1.615E+00 & 4.154E-02 & 3.353E+02 & 6.757E+01 & 4.402E-05 & 2.468E-06 \\
		130420A & 1.297  & 105   & -9.366E-01 & 1.738E-01 & -2.921E+00 & 4.152E-01 & 1.212E+02 & 8.541E+00 & 1.376E-05 & 5.858E-07 \\
		130427A & 0.3399 & 138.2 & -1.018E+00 & 1.843E-03 & -2.829E+00 & 3.238E-02 & 1.105E+03 & 7.299E+00 & 3.912E-03 & 2.169E-05 \\
		130518A & 2.488  & 48.58 & -8.629E-01 & 1.561E-02 & -2.181E+00 & 6.691E-02 & 1.329E+03 & 5.082E+01 & 1.280E-04 & 2.619E-06 \\
		130610A & 2.092  & 21.76 & -1.595E+00 & 7.599E-02 & -1.863E+01 & 2.230E+10 & 9.460E+02 & 4.954E+02 & 4.627E-06 & 3.145E-07 \\
		130702A & 0.145  & 58.88 & -1.736E-01 & 1.458E+01 & -2.453E+00 & 3.650E-02 & 1.110E+01 & 8.544E+00 & 9.512E-06 & 1.256E-05 \\
		130925A & 0.347  & 215.6 & -1.106E-01 & 1.831E-01 & -2.006E+00 & 3.260E-02 & 3.120E+01 & 1.089E+00 & 1.302E-04 & 3.620E-06 \\
		131011A & 1.874  & 77.06 & -8.778E-01 & 7.417E-02 & -2.085E+00 & 2.288E-01 & 6.255E+02 & 1.175E+02 & 1.227E-05 & 1.247E-06 \\
		131105A & 1.686  & 112.6 & -1.263E+00 & 2.244E-02 & -9.128E+00 & 5.123E+03 & 7.218E+02 & 4.918E+01 & 2.663E-05 & 2.127E-06 \\
		131108A & 2.4    & 18.18 & -9.140E-01 & 1.993E-02 & -2.464E+00 & 1.859E-01 & 1.247E+03 & 5.540E+01 & 4.496E-05 & 1.721E-06 \\
		131231A & 0.642  & 31.23 & -1.218E+00 & 9.639E-03 & -2.305E+00 & 3.373E-02 & 2.924E+02 & 6.619E+00 & 2.059E-04 & 2.734E-06 \\
		140206A & 2.73   & 27.26 & 5.490E-02  & 9.329E-02 & -2.416E+00 & 9.637E-02 & 4.521E+02 & 2.173E+01 & 1.780E-05 & 4.141E-07 \\
		140213A & 1.2076 & 18.62 & -1.126E+00 & 3.499E-02 & -2.252E+00 & 5.515E-02 & 1.902E+02 & 9.051E+00 & 2.799E-05 & 5.511E-07 \\
		140423A & 3.26   & 95.23 & -5.542E-01 & 1.155E-01 & -1.786E+00 & 4.965E-02 & 4.949E+02 & 6.768E+01 & 2.514E-05 & 6.316E-07 \\
		140506A & 0.889  & 64.13 & -1.180E+00 & 1.065E-01 & -1.682E+01 & 5.176E+07 & 3.715E+02 & 4.779E+01 & 7.174E-06 & 2.228E-06 \\
		140508A & 1.027  & 44.29 & -1.182E+00 & 1.890E-02 & -2.319E+00 & 9.382E-02 & 5.218E+02 & 2.457E+01 & 8.302E-05 & 2.704E-06 \\
		140512A & 0.725  & 148   & -1.225E+00 & 1.754E-02 & -3.540E+00 & 1.616E+01 & 1.192E+03 & 1.005E+02 & 4.073E-05 & 2.049E-05 \\
		140606B & 0.384  & 22.78 & -1.239E+00 & 4.621E-02 & -2.037E+00 & 4.716E-01 & 7.387E+02 & 1.561E+02 & 1.347E-05 & 3.904E-06 \\
		140620A & 2.04   & 45.83 & -8.497E-01 & 1.905E-01 & -2.092E+00 & 8.071E-02 & 2.112E+02 & 3.260E+01 & 8.075E-06 & 2.972E-07 \\
		140703A & 3.14   & 83.97 & -1.267E+00 & 5.914E-02 & -2.681E+00 & 9.169E-01 & 8.647E+02 & 1.438E+02 & 9.012E-06 & 7.920E-07 \\
		140801A & 1.32   & 7.168 & -3.846E-01 & 3.932E-02 & -3.853E+00 & 1.200E+00 & 2.770E+02 & 6.121E+00 & 1.279E-05 & 2.520E-07 \\
		140808A & 3.29   & 4.477 & -4.223E-01 & 1.004E-01 & -2.868E+00 & 4.827E-01 & 5.039E+02 & 2.771E+01 & 3.445E-06 & 1.376E-07 \\
		140907A & 1.21   & 35.84 & -1.018E+00 & 5.045E-02 & -1.373E+01 & 1.080E+06 & 3.082E+02 & 2.279E+01 & 6.932E-06 & 1.984E-06 \\
		141004A & 0.573  & 2.56  & 5.630E-02  & 5.375E-01 & -1.891E+00 & 7.405E-02 & 4.375E+01 & 1.045E+01 & 1.933E-06 & 1.533E-07 \\
		141028A & 2.33   & 31.49 & -8.420E-01 & 2.806E-02 & -1.966E+00 & 5.182E-02 & 9.760E+02 & 5.988E+01 & 5.019E-05 & 1.185E-06 \\
		141220A & 1.3195 & 7.616 & -8.324E-01 & 4.552E-02 & -1.027E+01 & 1.171E+04 & 4.153E+02 & 2.335E+01 & 5.550E-06 & 1.126E-06 \\
		141221A & 1.452  & 23.81 & -8.051E-01 & 2.786E-01 & -1.970E+00 & 2.347E-01 & 2.259E+02 & 7.044E+01 & 5.903E-06 & 8.685E-07 \\
		141225A & 0.915  & 56.32 & -2.982E-01 & 1.555E-01 & -2.059E+00 & 1.662E-01 & 3.415E+02 & 3.693E+01 & 5.638E-06 & 6.387E-07 \\
		150301B & 1.5169 & 13.31 & -1.049E+00 & 9.013E-02 & -2.222E+00 & 4.398E-01 & 4.606E+02 & 7.213E+01 & 4.089E-06 & 6.709E-07 \\
		150314A & 1.758  & 10.69 & -6.792E-01 & 1.382E-02 & -2.601E+00 & 1.021E-01 & 9.575E+02 & 2.178E+01 & 1.017E-04 & 2.333E-06 \\
		150403A & 2.06   & 22.27 & -8.733E-01 & 1.830E-02 & -2.108E+00 & 5.752E-02 & 1.312E+03 & 6.443E+01 & 7.922E-05 & 1.738E-06 \\
		150514A & 0.807  & 10.81 & -1.206E+00 & 9.780E-02 & -2.431E+00 & 1.824E-01 & 1.167E+02 & 1.067E+01 & 6.148E-06 & 2.786E-07 \\
		150727A & 0.313  & 49.41 & 1.314E-01  & 2.777E-01 & -2.158E+00 & 1.985E-01 & 1.947E+02 & 2.398E+01 & 6.557E-06 & 9.681E-07 \\
		150821A & 0.755  & 103.4 & -1.244E+00 & 1.676E-02 & -2.131E+00 & 9.393E-02 & 4.935E+02 & 3.002E+01 & 7.832E-05 & 3.769E-06 \\
		151027A & 0.81   & 123.4 & -1.247E+00 & 4.736E-02 & -1.955E+00 & 9.463E-02 & 3.645E+02 & 4.430E+01 & 2.287E-05 & 1.519E-06 \\
		160509A & 1.17   & 369.7 & -1.015E+00 & 9.913E-03 & -2.232E+00 & 4.756E-02 & 7.708E+02 & 2.143E+01 & 2.557E-04 & 4.928E-06 \\
		160625B & 1.406  & 453.4 & -9.341E-01 & 4.352E-03 & -2.182E+00 & 2.018E-02 & 1.134E+03 & 1.551E+01 & 9.629E-04 & 7.812E-06 \\
		160629A & 3.332  & 64.77 & -1.016E+00 & 3.634E-02 & -2.652E+00 & 1.504E+00 & 1.202E+03 & 9.250E+01 & 1.530E-05 & 2.664E-06 \\
		160804A & 0.736  & 131.6 & -1.030E+00 & 8.786E-02 & -2.819E+00 & 9.034E-01 & 1.239E+02 & 7.249E+00 & 1.907E-05 & 1.525E-06 \\
		161014A & 2.823  & 36.61 & -7.569E-01 & 8.587E-02 & -1.919E+01 & 4.129E+08 & 6.462E+02 & 5.513E+01 & 6.313E-06 & 6.033E-07 \\
		161017A & 2.0127 & 37.89 & -1.030E+00 & 1.010E-01 & -2.371E+00 & 7.765E-01 & 7.188E+02 & 1.228E+02 & 6.120E-06 & 1.209E-06 \\
		161117A & 1.549  & 122.2 & -8.111E-01 & 5.158E-02 & -3.023E+00 & 5.141E-01 & 2.056E+02 & 7.762E+00 & 3.479E-05 & 9.663E-07 \\
		161129A & 0.645  & 36.1  & -1.037E+00 & 9.784E-02 & -1.954E+00 & 1.622E-01 & 2.408E+02 & 7.009E+01 & 1.064E-05 & 1.388E-06 \\
		170214A & 2.53   & 122.9 & -9.788E-01 & 8.669E-03 & -2.512E+00 & 1.021E-01 & 1.699E+03 & 3.963E+01 & 2.283E-04 & 3.879E-06 \\
		170405A & 3.51   & 78.59 & -7.993E-01 & 1.996E-02 & -2.354E+00 & 8.888E-02 & 1.204E+03 & 4.189E+01 & 8.858E-05 & 1.487E-06 \\
		170607A & 0.557  & 20.93 & -1.286E+00 & 5.950E-02 & -2.383E+00 & 0.000E+00 & 1.741E+02 & 1.406E+01 & 1.243E-05 & 1.587E-07 \\
		170705A & 2.01   & 22.78 & -9.911E-01 & 6.923E-02 & -2.303E+00 & 1.083E-01 & 2.946E+02 & 2.301E+01 & 1.661E-05 & 4.933E-07 \\
		170728B & 1.272  & 46.34 & -9.615E-01 & 1.287E-01 & -2.436E+00 & 3.052E+00 & 3.101E+02 & 6.521E+01 & 4.927E-06 & 4.034E-06 \\
		170903A & 0.886  & 25.6  & -1.312E+00 & 1.088E-01 & -2.434E+01 & 7.713E+11 & 1.793E+02 & 2.525E+01 & 4.096E-06 & 4.273E-06 \\
		171010A & 0.33   & 107.3 & -1.089E+00 & 5.936E-03 & -2.191E+00 & 8.671E-03 & 1.831E+02 & 1.898E+00 & 9.013E-04 & 4.507E-06 \\
		180205A & 1.409  & 15.36 & -7.653E-01 & 7.705E-01 & -2.130E+00 & 1.681E-01 & 8.480E+01 & 4.100E+01 & 2.790E-06 & 3.265E-07 \\
		180314A & 1.445  & 22.02 & -4.038E-01 & 7.725E-02 & -3.356E+00 & 1.473E+00 & 2.517E+02 & 1.097E+01 & 1.399E-05 & 7.652E-07 \\
		180620B & 1.1175 & 46.72 & -1.206E+00 & 1.156E-01 & -1.660E+00 & 3.474E-02 & 3.719E+02 & 1.054E+02 & 1.809E-05 & 6.632E-07 \\
		180703A & 0.6678 & 20.74 & -7.756E-01 & 4.122E-02 & -1.967E+00 & 1.032E-01 & 5.850E+02 & 5.378E+01 & 2.873E-05 & 2.242E-06 \\
		180720B & 0.654  & 48.9  & -1.171E+00 & 4.805E-03 & -2.490E+00 & 7.095E-02 & 1.052E+03 & 2.551E+01 & 4.490E-04 & 8.034E-06 \\
		180728A\tnote{$a$} & 0.117  & 6.4   & -1.540E+00 & 1.000E-02 & -2.460E+00 & 2.000E-02 & 8.847E+01 & 1.564E+00 & 7.835E-05 & 4.395E-07 \\
		181020A\tnote{$a$} & 2.938  & 15.1  & -7.000E-01 & 2.000E-02 & -2.060E+00 & 7.000E-02 & 1.445E+03 & 6.695E+01 & 3.840E-05 & 8.864E-07 \\
		190114C\tnote{$a$} & 0.425  & 116.4 & -1.058E+00 & 3.000E-03 & -3.180E+00 & 7.000E-02 & 1.423E+03 & 1.696E+01 & 7.371E-04 & 5.482E-06 \\
		190324A\tnote{$a$} & 1.1715 & 26.88 & -8.400E-01 & 4.000E-02 & -2.060E+00 & 5.000E-02 & 3.133E+02 & 1.824E+01 & 2.348E-05 & 7.454E-07 \\
		190719C\tnote{$b$} & 2.469  & 175.6 & -8.600E-01 & 2.100E-01 & -2.740E+00 & 5.800E-01 & 2.780E+02 & 4.059E+01 & 8.098E-06 & 4.550E-07 \\
		190829A\tnote{$a$} & 0.0785 & 59.39 & -9.200E-01 & 6.200E-01 & -2.510E+00 & 1.000E-02 & 1.186E+01 & 1.079E+00 & 2.681E-05 & 2.428E-06 \\
		200524A\tnote{$a$} & 1.256  & 37.76 & -6.600E-01 & 6.000E-02 & -1.770E+00 & 3.000E-02 & 4.332E+02 & 3.610E+01 & 2.502E-05 & 6.818E-07 \\
		200613A\tnote{$a$} & 1.228  & 478   & -1.080E+00 & 2.000E-02 & -2.580E+00 & 8.000E-02 & 2.473E+02 & 6.684E+00 & 5.873E-05 & 8.958E-07 \\
		201020B\tnote{$a$} & 0.804  & 15.87 & -7.100E-01 & 2.000E-02 & -2.200E+00 & 3.000E-02 & 2.470E+02 & 6.134E+00 & 4.630E-05 & 7.386E-07 \\
		201216C\tnote{$a$} & 1.1    & 29.95 & -1.060E+00 & 1.000E-02 & -2.250E+00 & 3.000E-02 & 6.846E+02 & 1.470E+01 & 1.921E-04 & 2.315E-06 \\
		210104A\tnote{$b$} & 0.46   & 31.75 & -1.101E+00 & 7.550E-02 & -2.449E+00 & 4.020E-01 & 2.808E+02 & 3.913E+01 & 2.467E-05 & 3.315E-06 \\
		210204A\tnote{$a$} & 0.876  & 206.9 & -1.500E+00 & 1.000E-01 & -2.300E+00 & 5.000E-01 & 2.626E+02 & 9.380E+01 & 1.068E-04 & 1.592E-05 \\
		210610A\tnote{$b$} & 3.54   & 8.192 & -4.720E-01 & 5.670E-01 & -1.983E+00 & 1.770E-01 & 3.154E+02 & 1.130E+02 & 1.604E-06 & 1.124E-07 \\
		210610B\tnote{$b$} & 1.13   & 55.04 & -3.700E-01 & 2.000E-02 & -3.080E+00 & 2.300E-01 & 7.977E+02 & 1.815E+01 & 1.302E-04 & 4.531E-06 \\
		210619B\tnote{$a$} & 1.937  & 54.79 & -8.600E-01 & 1.000E-02 & -1.990E+00 & 1.000E-02 & 6.168E+02 & 8.811E+00 & 4.341E-04 & 2.213E-06 \\
		210722A\tnote{$b$} & 1.145  & 61.95 & -8.770E-01 & 2.830E-01 & -3.407E+00 & 5.420E+00 & 2.390E+02 & 6.435E+01 & 4.756E-06 & 8.564E-07 \\
		210731A\tnote{$b$} & 1.2525 & 25.86 & 5.000E-02  & 1.900E-01 & -3.300E+00 & 1.170E+00 & 4.343E+02 & 4.009E+01 & 3.236E-06 & 3.190E-07 \\
		211023A\tnote{$a$} & 0.39   & 79.11 & -1.740E+00 & 1.000E-02 & -2.550E+00 & 7.000E-02 & 1.279E+02 & 2.780E+00 & 1.367E-04 & 1.530E-06 \\
		220101A\tnote{$a$} & 4.618  & 128.3 & -1.060E+00 & 2.000E-02 & -2.300E+00 & 2.000E-01 & 1.629E+03 & 1.011E+02 & 7.116E-05 & 1.810E-06 \\
		220107A\tnote{$a$} & 1.246  & 33.03 & -5.500E-01 & 6.000E-02 & -1.940E+00 & 6.000E-02 & 3.773E+02 & 2.471E+01 & 2.811E-05 & 1.202E-06 \\
		220521A\tnote{$b$} & 5.6    & 13.57 & -1.600E-01 & 7.900E-01 & -2.910E+00 & 6.600E-01 & 3.410E+02 & 6.600E+01 & 9.846E-07 & 6.278E-08 \\
		220527A\tnote{$a$} & 0.857  & 10.5  & -7.500E-01 & 2.000E-02 & -2.550E+00 & 5.000E-02 & 2.815E+02 & 5.014E+00 & 6.180E-05 & 8.231E-07 \\
		220627A\tnote{$a$} & 3.084  & 136.7 & -8.100E-01 & 3.000E-02 & -2.680E+00 & 3.100E-01 & 1.593E+03 & 8.168E+01 & 5.431E-05 & 2.216E-06 \\
		221226B\tnote{$b$} & 2.694  & 4.864 & 2.600E-01  & 3.100E-01 & -5.180E+00 & 1.580E+01 & 3.794E+02 & 3.642E+01 & 1.153E-06 & 2.311E-08 \\
		230204B\tnote{$a$} & 2.142  & 216.1 & -9.700E-01 & 2.000E-02 & -2.730E+00 & 2.900E-01 & 2.397E+03 & 1.414E+02 & 1.020E-04 & 3.063E-06 \\
		230812B\tnote{$a$} & 0.36   & 3.264 & -8.000E-01 & 1.000E-02 & -2.470E+00 & 2.000E-02 & 3.713E+02 & 4.080E+00 & 3.702E-04 & 2.923E-06 \\
		230818A\tnote{$b$} & 2.42   & 9.981 & -8.600E-01 & 7.000E-02 & -2.510E+00 & 4.500E-01 & 7.312E+02 & 8.926E+01 & 4.387E-06 & 3.846E-07 \\\hline
	\end{longtable}
\end{ThreePartTable}

\newpage

\begin{ThreePartTable}\scriptsize
	\begin{TableNotes}
		\item[$a$] The spectral parameters of 1 GRBs are taken from the GCN.
		\item[$b$] The spectral parameters of 5 GRBs are obtained in this work.
	\end{TableNotes}
	\begin{longtable}{lllllllllcc}
		\caption{The 28 LGRB data in the FULL sample which are not in the GOLD sample.}\label{tab:addLGRBs} \\
		\hline\hline
		GRB & $z$ & $T_{90}$ & $\alpha$ & $\sigma_{\alpha}$ & $\beta$ & $\sigma_{\beta}$ & $E_{\rm p,i}(\rm keV)$ & $\sigma_{E_{\rm p,i}}(\rm keV)$ & $S_{\rm bolo}(\rm erg/cm^2)$ & $\sigma_{S_{\rm bolo}}(\rm erg/cm^2)$ \\
		\hline
		\endfirsthead 
		\caption*{The 28 additional LGRB data in the FULL sample compared to the GOLD sample. (continued)}\label{tab:addLGRBs} \\
		\hline\hline
		GRB & $z$ & $T_{90}$ & $\alpha$ & $\sigma_{\alpha}$ & $\beta$ & $\sigma_{\beta}$ & $E_{\rm p,i}(\rm keV)$ & $\sigma_{E_{\rm p,i}}(\rm keV)$ & $S_{\rm bolo}(\rm erg/cm^2)$ & $\sigma_{S_{\rm bolo}}(\rm erg/cm^2)$ \\
		\hline
		\endhead 
		\hline
		\endfoot 
		\insertTableNotes
		\endlastfoot 
		080905B & 2.374  & 106   & -8.523E-01 & 1.736E-01 & -2.316E+00 & 1.645E+01 & 6.160E+02 & 1.739E+02 & 3.578E-06 & 1.564E-05 \\
		080916C & 4.35   & 62.98 & -1.070E+00 & 1.449E-02 & -2.152E+00 & 7.649E-02 & 3.572E+03 & 2.230E+02 & 7.770E-05 & 8.500E-07 \\
		081109A & 0.9787 & 58.37 & -1.899E-01 & 2.731E-01 & -1.649E+00 & 5.174E-02 & 5.661E+01 & 2.173E+01 & 1.314E-05 & 8.112E-07 \\
		090423  & 8.2    & 7.168 & -5.529E-01 & 3.648E-01 & -2.620E+00 & 1.287E+00 & 6.097E+02 & 9.979E+01 & 8.734E-07 & 8.334E-08 \\
		101219B & 0.5519 & 51.01 & 8.853E-01  & 4.932E-01 & -2.391E+00 & 2.761E-01 & 9.825E+01 & 2.501E+01 & 4.734E-06 & 5.468E-07 \\
		110106B & 0.618  & 35.52 & -9.437E-01 & 1.370E-01 & -8.148E+00 & 3.141E+03 & 2.121E+02 & 2.408E+01 & 4.385E-06 & 1.891E-06 \\
		110818A & 3.36   & 67.07 & -1.054E+00 & 1.357E-01 & -1.805E+00 & 1.686E-01 & 8.056E+02 & 2.223E+02 & 7.186E-06 & 5.006E-07 \\
		111107A & 2.893  & 12.03 & -1.252E+00 & 1.739E-01 & -2.000E+00 & 0.000E+00 & 8.326E+02 & 3.781E+02 & 1.233E-06 & 5.329E-08 \\
		120118B & 2.943  & 37.83 & -2.970E-01 & 2.542E-01 & -2.547E+00 & 1.492E-01 & 1.698E+02 & 1.206E+01 & 3.020E-06 & 9.164E-08 \\
		120922A & 3.1    & 182.3 & -5.366E-01 & 1.096E+00 & -2.228E+00 & 7.979E-01 & 1.476E+02 & 1.665E+01 & 1.002E-05 & 1.723E-06 \\
		121211A & 1.023  & 5.632 & -2.936E-01 & 2.445E-01 & -4.900E+00 & 1.867E+01 & 2.048E+02 & 2.803E+01 & 6.602E-07 & 4.518E-08 \\
		130612A & 2.006  & 7.424 & 3.295E-01  & 5.639E-01 & -2.254E+00 & 1.824E-01 & 8.699E+01 & 2.475E+01 & 8.187E-07 & 8.919E-08 \\
		131229A & 1.04   & 12.99 & -7.258E-01 & 2.305E-02 & -4.309E+00 & 3.085E+00 & 7.728E+02 & 2.710E+01 & 2.890E-05 & 2.238E-06 \\
		140304A & 5.283  & 31.23 & -7.893E-01 & 1.765E-01 & -2.429E+00 & 6.779E-01 & 7.692E+02 & 1.975E+02 & 2.691E-06 & 1.640E-07 \\
		140623A & 1.92   & 111.1 & -1.422E+00 & 9.893E-02 & -1.630E+01 & 1.163E+08 & 9.535E+02 & 4.037E+02 & 3.873E-06 & 2.532E-07 \\
		140713A & 0.935  & 5.376 & 2.956E-01  & 7.273E-01 & -1.806E+00 & 1.272E-01 & 7.130E+01 & 2.571E+01 & 1.409E-06 & 1.946E-07 \\
		151111A & 3.5    & 46.34 & -7.834E-02 & 2.810E-01 & -1.327E+01 & 1.421E+05 & 5.339E+02 & 2.265E+02 & 2.239E-06 & 2.228E-07 \\
		151229A & 1.4    & 3.456 & -1.307E+00 & 9.063E-02 & -1.042E+01 & 9.012E+04 & 2.496E+02 & 3.396E+01 & 1.303E-06 & 6.560E-07 \\
		161001A & 0.67   & 2.24  & -9.395E-01 & 8.287E-02 & -4.363E+00 & 2.746E+01 & 6.223E+02 & 9.954E+01 & 1.957E-06 & 1.054E-06 \\
		170113A & 1.968  & 49.15 & -1.704E+00 & 2.232E-01 & -2.046E+01 & 2.384E+10 & 3.339E+02 & 1.745E+02 & 2.861E-06 & 4.689E-07 \\
		171222A & 2.409  & 80.38 & -3.713E-01 & 1.114E+00 & -2.135E+00 & 9.542E-02 & 5.980E+01 & 1.410E+01 & 4.325E-06 & 4.636E-07 \\
		180418A & 1.55   & 2.56  & -1.389E+00 & 9.523E-02 & -9.407E+00 & 5.234E+04 & 2.680E+03 & 2.426E+03 & 9.243E-07 & 2.409E-07 \\
		181010A\tnote{$b$} & 1.39   & 9.728 & -6.000E-02 & 1.430E+00 & -1.800E+00 & 1.700E-01 & 1.423E+02 & 9.536E+01 & 1.260E-06 & 2.235E-07 \\
		190613A\tnote{$b$} & 2.78   & 17.15 & -9.190E-01 & 1.680E-01 & -3.013E+00 & 1.100E+00 & 4.502E+02 & 6.124E+01 & 3.542E-06 & 2.371E-07 \\
		191011A\tnote{$b$} & 1.722  & 25.09 & -1.130E+00 & 5.200E-01 & -2.910E+00 & 1.730E+00 & 1.502E+02 & 5.063E+01 & 1.050E-06 & 1.645E-07 \\
		200829A\tnote{$a$} & 1.25   & 6.912 & -4.300E-01 & 1.000E-02 & -2.400E+00 & 2.000E-02 & 7.578E+02 & 9.675E+00 & 2.891E-04 & 2.192E-06 \\
		201020A\tnote{$b$} & 2.903  & 21.5  & -9.400E-01 & 3.500E-01 & -2.470E+00 & 2.100E-01 & 1.332E+02 & 1.975E+01 & 1.993E-06 & 1.528E-07 \\
		201021C\tnote{$b$} & 1.07   & 35.33 & -8.319E-01 & 3.200E-01 & -2.308E+00 & 6.820E-01 & 2.575E+02 & 9.108E+01 & 2.134E-06 & 5.552E-07 \\ \hline
	\end{longtable}
\end{ThreePartTable}
\twocolumn


\begin{thebibliography}{}
	\makeatletter
	\relax
	\def\mn@urlcharsother{\let\do\@makeother \do\$\do\&\do\#\do\^\do\_\do\%\do\~}
	\def\mn@doi{\begingroup\mn@urlcharsother \@ifnextchar [ {\mn@doi@}
		{\mn@doi@[]}}
	\def\mn@doi@[#1]#2{\def\@tempa{#1}\ifx\@tempa\@empty \href
		{http://dx.doi.org/#2} {doi:#2}\else \href {http://dx.doi.org/#2} {#1}\fi
		\endgroup}
	\def\mn@eprint#1#2{\mn@eprint@#1:#2::\@nil}
	\def\mn@eprint@arXiv#1{\href {http://arxiv.org/abs/#1} {{\tt arXiv:#1}}}
	\def\mn@eprint@dblp#1{\href {http://dblp.uni-trier.de/rec/bibtex/#1.xml}
		{dblp:#1}}
	\def\mn@eprint@#1:#2:#3:#4\@nil{\def\@tempa {#1}\def\@tempb {#2}\def\@tempc
		{#3}\ifx \@tempc \@empty \let \@tempc \@tempb \let \@tempb \@tempa \fi \ifx
		\@tempb \@empty \def\@tempb {arXiv}\fi \@ifundefined
		{mn@eprint@\@tempb}{\@tempb:\@tempc}{\expandafter \expandafter \csname
			mn@eprint@\@tempb\endcsname \expandafter{\@tempc}}}
	
	
	\bibitem[Akritas \& Bershady(1996)]{akritas_linear_1996} Akritas M.~G.,\& Bershady M.~A. \href{https://doi.org/10.1086/177901}{1996, ApJ, 470, 706}
	
	\bibitem[Amati et al.(2006)]{Amati_2006} Amati L. \href{https://doi.org/10.1111/j.1365-2966.2006.10840.x}{2006, MNRAS, 372, 233}
	
	\bibitem[Amati et al.(2009)]{Amati2009} Amati, L., Frontera, F., \& Guidorzi, C. \href{https://doi.org/10.1051/0004-6361/200912788}{2009, A\&A, 508, 173}
	
	\bibitem[Amati \&  Valle(2013)]{AMATI_2013} Amati L., \& Valle M.~D. \href{https://doi.org/10.1142/s0218271813300280}{2013, IJMPD, 22, 1330028}
	\bibitem[Amati et al.(2002)]{Amati_2002} Amati L., Frontera F., Tavani M., et al.,
	\href{https://doi.org/10.1051/0004-6361:20020722}{2002 , A\&A, 390, 81}
	
	\bibitem[Amati et al.(2008)]{Amati_2008} Amati, Guidorzi, Frontera, et al., \href{https://doi.org/10.1111/j.1365-2966.2008.13943.x}{2008, MNRAS, 391, 577}
	
	\bibitem[Amati \& Della Valle.(2013)]{Amati2013} Amati L., Della Valle M., \href{https://doi.org/10.1142/S0218271813300280}{2013, IJMPD, 22, 1330028}

	\bibitem[Amati et al.(2018)]{Amati_2018} Amati L., O'Brien P., Gotz D., et al., \href{https://doi.org/10.1016/j.asr.2018.03.010}{2018, AdSpR, 62, 191}
	
	\bibitem[Amati et al.(2019)]{Amati_2019} Amati L., D'Agostino R., Luongo O., et al. \href{https://doi.org/10.1093/mnrasl/slz056}{2019, MNRAS, 486, L46}
	
	
	\bibitem[Atteia et al.(2017)]{Atteia2017} Atteia J.-L.,  Heussaff, V., Dezalay, J.-P.  et al.,
	\href{https://doi.org/10.3847/1538-4357/aa5ffa}{2017, ApJ, 837, 119}
	
	\bibitem[Band et al.(1993)]{Band_1993} Band D., Matteson J., Ford L., et al.,
	\href{https://doi.org/10.1063/1.44238}{1993, ApJ, 280, 872}
	
	
	\bibitem[Bernardini et al.(2012)]{Bernardini_2012} Bernardini M.~G., Margutti R., Zaninoni E., \& Chincarini G. \href{https://doi.org/10.1111/j.1365-2966.2012.21487.x}{2012, MNRAS, 425, 1199}
	
	\bibitem[Bhat et al.(2016)]{Bhat_2016} Narayana Bhat P., Meegan C.~A., von Kienlin A., et al., \href{https://doi.org/10.3847/0067-0049/223/2/28}{2016, ApJS, 223, 28}
	
	\bibitem[Brout et al.(2022)]{Brout2022} Brout D., et al.,
	\href{https://doi.org/10.3847/1538-4357/ac8e04}{2022, ApJ, 938, 110}
	
	\bibitem[Butler et al.(2007)]{Butler2007} Butler N.~R., Kocevski D., Bloom J.~S., Curtis J.~L., \href{https://doi.org/10.1086/522492}{2007, ApJ, 671, 656}
	\bibitem[Butler, Kocevski, \& Bloom(2009)]{Butler2009} Butler N.~R., Kocevski D., Bloom J.~S., \href{https://doi.org/10.1088/0004-637X/694/1/76}{2009, ApJ, 694, 76}
	\bibitem[Butler, Bloom, \& Poznanski(2010)]{Butler2010} Butler N.~R., Bloom J.~S., Poznanski D., \href{https://doi.org/10.1088/0004-637X/711/1/495}{2010, ApJ, 711, 495}
	
	\bibitem[Chevallier \& Polarski(2001)]{CP2001} Chevallier, M. \& Polarski, D.
	\href{https://doi.org/10.1142/S0218271801000822}{2001, IJMPD, 10, 213}
	\bibitem[Cucchiara et al.(2011)]{Cucchiara2011} Cucchiara, A., Levan, A., Fox, D. B., et al., \href{https://doi.org/10.1088/0004-637X/736/1/7}{2011, ApJ, 736, 7}
	
	\bibitem[Cunningham \& SEDM Team (2019)]{Cunningham_2019} Cunningham V., \& SEDM Team, \href{https://ui.adsabs.harvard.edu/abs/2019GCN.24835....1C}{2019, GCN, 24835}
	
	\bibitem[D'Agostini(2005)]{dagostini2005fits}D'Agostini G. \href{https://arxiv.org/abs/physics/0511182v1}{2005, arXiv: physics/0511182}
	
	\bibitem[Dai et al.(2004)]{Dai2004} Dai, Z., Liang, E., \& Xu, D.
	\href{https://doi.org/10.1086/424694}{2004, ApJ, 612, L101}
	
	\bibitem[Dainotti \& Amati(2018)]{Dainotti_2018} Dainotti M. G., \& Amati L.
	\href{https://doi.org/10.1088/1538-3873/aaa8d7}{2018, PASP, 130, 051001}
	
	\bibitem[Dainotti \& Del~Vecchio(2017)]{Dainotti_2017} Dainotti M., \& Del~Vecchio R. \href{https://doi.org/10.1016/j.newar.2017.04.001}{2017, New Astronomy Reviews, 77, 23}
	
	\bibitem[Dainotti et al.(2008)]{Dainotti_2008} Dainotti M.~G., Cardone V.~F., \& Capozziello S.
	\href{https://doi.org/10. 111/j.1745-3933.2008.00560.x}{2008, MNRAS, 391, L79}
	
	\bibitem[Dainotti et al.(2011)]{Dainotti_2011} Dainotti M.~G., Ostrowski M., \& Willingale R. \href{https://doi.org/10.1111/j.1365-2966.2011.19433.x}{2011, MNRAS, 418, 2202}
	
	\bibitem[Dainotti et al.(2017)]{Dainotti_2016b} Dainotti M.~G., Nagataki S., Maeda K., et al., \href{https://doi.org/10.1051/0004-6361/201628384}{2017, A\&A, 600, A98}
	

	\bibitem[Dainotti et al.(2023)]{Dainotti2023} Dainotti M.~G., Levine D., Fraija N., et al.,
	\href{https://doi.org/10.3390/galaxies11010025}{2023, Galaxies, 11, 25}
	
	\bibitem[Demianski et al.(2017a)]{Demianski_2017a} Demianski M., Piedipalumbo E., Sawant D., et al., \href{https://doi.org/10.1051/0004-6361/201628909}{2017, A\&A, 598, A112}
	
	\bibitem[Demianski et al.(2017b)]{Demianski_2017b} Demianski M., Piedipalumbo E., Sawant D., et al., \href{https://doi.org/10.1051/0004-6361/201628911}{2017, A\&A, 598, A113}
	
	\bibitem[Dirirsa et al.(2019)]{Dirirsa2019} Fana Dirirsa F., Razzaque S., Piron F., et al., \href{https://doi.org/10.3847/1538-4357/ab4e11}{2019, ApJ, 887, 13}
	
	\bibitem[Fenimore \& Ramirez-Ruiz(2000)]{fenimore_2000} Fenimore E.~E., \& Ramirez-Ruiz E. \href{https://ui.adsabs.harvard.edu/abs/2000astro.ph..4176F}{arXiv: astro-ph/0004176}
	
	\bibitem[Foreman-Mackey et al.(2013)]{Foreman-Mackey_2013}Foreman-Mackey D., Conley A., Meierjurgen Farr W., et al., \href{https://doi.org/10.1086/670067}{,125, 306}
	\bibitem[Frontera et al.(2012)]{Frontera2012} Frontera F., Amati L., Guidorzi C., et al., \href{https://doi.org/10.1088/0004-637X/754/2/138}{2012, ApJ, 754, 138}
	
	\bibitem[Ghirlanda et al.(2004)]{Ghirlanda_2004} Ghirlanda G., Ghisellini G.,\& Lazzati D. \href{https://doi.org/10.1086/424913}{2004, ApJ, 616, 331}
	
	\bibitem[Ghirlanda et al.(2006)]{Ghirlanda2006} Ghirlanda, G., Ghisellini, G.,\& Firmani, C. \href{https://doi.org/10.1088/1367-2630/8/7/123}{2006, New, J. Phys., 8, 123}
	
	\bibitem[Ghirlanda et al.(2007)]{Ghirlanda_2007} Ghirlanda G., Nava L., Ghisellini G., et al., \href{https://doi.org/10.1051/0004-6361:20077119}{2007, A\&A, 466, 127}
	
	\bibitem[Goldstein et al.(2017)]{Goldstein_2017}  Goldstein A., Veres P., Burns E., Briggs M.~S., et al., \href{https://doi.org/10.3847/2041-8213/aa8f41}{2017, ApJ, 848, L14}
	
	
	\bibitem[Gehrels et al.(2004)]{Gehrels2004} Gehrels N., Chincarini G., Giommi P., et al., \href{https://doi.org/10.1086/422091}{2004, ApJ, 611, 1005}
	\bibitem[Ghirlanda et al.(2008)]{Ghirlanda2008} Ghirlanda G., Nava L., Ghisellini G., et al., \href{https://doi.org/10.1111/j.1365-2966.2008.13232.x}{2008, MNRAS, 387, 319}
	
	\bibitem[Goldstein et al.(2012)]{Goldstein2012} Goldstein A., Burgess J.~M., Preece R.~D., et al., \href{https://doi.org/10.1088/0067-0049/199/1/19}{2012, ApJS, 199, 19}
	
	\bibitem[Gruber et al.(2014)]{gruber2014fermi} Gruber D., Goldstein A., Weller von Ahlefeld V., et al., \href{https://doi.org/10.1088/0067-0049/211/1/12}{2014, ApJS, 211, 12}
	
	\bibitem[Hogg, Bovy, \& Lang(2010)]{Hogg_2010} Hogg D.~W., Bovy J., \& Lang D., \href{https://doi.org/10.48550/arXiv.1008.4686}{2010, arXiv, arXiv:1008.4686.}
	
	\bibitem[Heintz et al.(2019)]{Heintz_2019} Heintz, K.~E., Fynbo J.~P.~U., de Ugarte Postigo A., \href{https://ui.adsabs.harvard.edu/abs/2019GCN.24686....1H}{2019, GCN, 24686}
	
	\bibitem[Higgins et al. (2019)]{Higgins2019}  Higgins, A. B., van der Horst, A. J., Starling, R. L. C. et al.
	\href{https://doi.org/10.1093/mnras/stz384}{2019, MNRAS, 484,  5245}
	
	
	
	
	
	\bibitem[Iyyani et al.(2013)]{Iyyani_2013}  Iyyani, S., Ryde F., Axelsson, M., et al., \href{https://doi.org/10.1093/mnras/stt863}{2013, MNRAS, 433, 2739}
	
	\bibitem[Izzo et al.(2015)]{Izzo_2015} Izzo, L.,  Muccino, M.,  Zaninoni, E.,  Amati L., et al., \href{https://doi.org/10.1051/0004-6361/201526461}{2015, A \& A, 582, A115}
	
	\bibitem[Jia et al.(2022)]{Jia_2022} Jia X.~D., Hu J.~P., Yang J., Zhang B.~B., Wang F.~Y. \href{https://doi.org/10.1093/mnras/stac2356}{2022, MNRAS, 516, 2575}
	
	\bibitem[Jordana-Mitjans et al.(2022)]{Jordana2022} Jordana-Mitjans N., Mundell C.~G., Guidorzi C., et al., \href{https://doi.org/10.3847/1538-4357/ac972b}{2022, ApJ, 939, 106}
	
	\bibitem[Khadka \& Ratra(2020)]{Khadka2020} Khadka, N. \& Ratra, B.
	\href{https://doi.org/10.1093/mnras/staa2779}{2020, MNRAS, 499, 391 }
	
	\bibitem[Khadka et al.(2021)]{Khadka2021} Khadka, N., Luongo, O., Muccino, M., \& Ratra, B. \href{https://doi.org/10.1088/1475-7516/2021/09/042}{2021, JCAP, 09, 042}
	
	\bibitem[Li, Zhang \& Liang(2023)]{Li_2023} Li Z., Zhang B., Liang N. \href{https://doi.org/10.1093/mnras/stad838}{2023, MNRAS, 521, 4406}
	
	\bibitem[Liang \& Zhang(2005)]{Liang_2005} Liang E., Zhang B. \href{https://doi.org/10.1086/491594}{2005, ApJ, 633, 611}
	\bibitem[Liang \& Zhang(2006)]{Liang2006} Liang, E., \& Zhang, B.
	\href{https://doi.org/10.1111/j.1745-3933.2006.00169.x}{2006, MNRAS, 369, L37}
	
	\bibitem[Liang, Xiao, Liu \& Zhang(2008)]{Liang_2008} Liang N., Xiao W.~K., Liu Y. \href{https://doi.org/10.1086/590903}{2008, ApJ, 685, 354}
	\bibitem[Li, Zhang \& Liang(2023)]{LZL2023} Li, Z., Zhang, B.,  \& Liang, N.
	\href{https://doi.org/10.1093/mnras/stad838}{ 2023, MNRAS, 521, 4406}
	\bibitem[Li et al.(2023)]{Li2023} Li, J.-L., Yang, Y.-P., Yi, S.-X., Hu, J.-P., Wang, F.-Y., \& Qu, Y.-K.
	\href{https://doi.org/10.3847/1538-4357/ace107}{2023, ApJ, 953, 58}
	\bibitem[Liang et al.(2008)]{Liang2008} Liang, N., Xiao, W. K., Liu, Y., \& Zhang, S. N. \href{https://doi.org/10.1086/590903}{2008, ApJ, 685, 354}
	\bibitem[Liang \& Zhang(2008)]{LiangZhang2008} Liang, N., \& Zhang, S.
	\href{https://doi.org/10.1063/1.3027949}{2008, AIP Conf. Proc. Vol. 1065, Am. Inst. Phys New York}
	\bibitem[Liang et al.(2010)]{Liang2010} Liang, N., Wu, P.,  \& Zhang, S. N.
	\href{https://doi.org/10.1103/PhysRevD.81.083518}{2010, PRD, 81, 083518}
	\bibitem[Liang et al.(2011)]{Liang2011}  Liang, N., Xu, L.,  \& Zhu, Z. H.
	\href{https://doi.org/10.1051/0004-6361/201015919}{2011, A\&A, 527, A11}
	\bibitem[Liang et al.(2022)]{Liang2022} Liang, N., Li, Z., Xie, X., \& Wu, P.  \href{https://doi.org/10.3847/1538-4357/aca08a}{2022, ApJ, 941, 84}
	\bibitem[Linder(2003)]{L2003} Linder, E.V.
	\href{https://doi.org/10.1103/PhysRevLett.90.091301}{2003, PRL,  90, 091301}
	\bibitem[Lin et al.(2015)]{Lin_2015} Lin H.-N., Li X., Chang Z.
	\href{https://doi.org/10.1093/mnras/stv2471}{2015, MNRAS, 455, 2131}
	\bibitem[Liu \& Wei(2015)]{Liu2015} Liu, J., \& Wei, H.
	\href{https://doi.org/10.1007/s10714-015-1986-1}{2015, GReGr, 47, 141}
	\bibitem[Liu et al.(2022a)]{Liu2022a} Liu, Y., Chen, F., Liang, N., et al.
	\href{https://doi.org/10.3847/1538-4357/ac66d3}{2022, ApJ, 931, 50}
	\bibitem[Liu et al.(2022b)]{Liu2022b} Liu, Y., Liang, N., Xie, X., et al.
	\href{https://doi.org/10.3847/1538-4357/ac7de5}{2022, ApJ, 935, 7}
	\bibitem[Luongo \& Muccino(2021a)]{luongo2021roadmap} Luongo O., Muccino M. \href{https://ui.adsabs.harvard.edu/abs/2021Galax...9...77L}{2021a, Galaxy, 9, 77}
	\bibitem[Luongo \& Muccino(2021b)]{Luongo2021b} Luongo, O., \& Muccino, M.
	\href{https://doi.org/10.1093/mnras/stab795}{2021b, MNRAS, 503, 4581}
	\bibitem[Luongo \& Muccino(2023)]{Luongo2023} Luongo, O., \& Muccino, M.
	\href{https://doi.org/10.1093/mnras/stac2925}{2023, MNRAS, 518, 2247}
	
	\bibitem[Lusso et al.(2019)]{Lusso_2019} Lusso E., Piedipalumbo E., Risaliti G., \& Paolillo M. \href{https://doi.org/10.1051/0004-6361/201936223}{2019, A \& A, 628, L4}
	
	\bibitem[Margutti et al.(2012)]{Margutti_2012} Margutti R., Zaninoni E., Bernardini M.~G., et al., \href{https://doi.org/10.1093/mnras/sts066}{2012, MNRAS, 428, 729}
	
	\bibitem[Meegan et al.(2009)]{Meegan_2009} Meegan C., Lichti G., Bhat P.~N., et al., \href{https://doi.org/10.1088/0004-637x/702/1/791}{2009, ApJ, 702, 791}
	
	\bibitem[Montiel et al.(2021)]{Montiel2021} Montiel, A., Cabrera, J. I., \&  Hidalgo, J. C. \href{https://doi.org/10.1093/mnras/staa3926}{2021, MNRAS, 501, 3515}
	
	\bibitem[Moresco et al.(2012)]{Moresco_2012} Moresco, M., Verde, L., Pozzetti, L., Jimenez, R. \& Cimatti, A. \href{https://doi.org/10.1088/1475-7516/2012/08/006}{2012, JCAP, 08, 006}
	\bibitem[Moresco (2015)]{Moresco_2015} Moresco, M.
	\href{https://doi.org/10.1093/mnrasl/slv037}{2015, MNRAS, 450, L16}
	\bibitem[Moresco et al.(2016)]{Moresco_2016} Moresco, M., Pozzetti, L., Cimatti, A. et al. \href{https://doi.org/10.1088/1475-7516/2016/05/014}{2016, JCAP, 05, 014}
	\bibitem[Moresco et al.(2020)]{Moresco_2020} Moresco M., Jimenez R., Verde L., et al.
	\href{https://doi.org/10.3847/1538-4357/ab9eb0 }{2020, ApJ, 898, 82}
	\bibitem[Moresco et al.(2022)]{Moresco_2022} Moresco M., Amati L., Amendola L., et al., \href{https://doi.org/10.1007/s41114-022-00040-z}{2022, LRR, 25, 1}
	
	\bibitem[Nava et al.(2012b)]{Nava2012} Nava L., Salvaterra R., Ghirlanda G., et al., \href{https://doi.org/10.1111/j.1365-2966.2011.20394.x}{2012, MNRAS, 421, 1256}
	
	\bibitem[Nemmen et al.(2012)]{nemmen_universal_2012} Nemmen R.~S., Georganopoulos M.,  Guiriec S., et al, \href{https://doi.org/10.1126/science.1227416}{2012, Science, 338, 1445}
	
	\bibitem[Norris, Marani, \& Bonnell(2000)]{Norris_2000} Norris J.~P., Marani G.~F., \& Bonnell J.~T. \href{https://doi.org/10.1086/308725}{2000, ApJ, 534, 248}
	
	Oates, S. Universe 2023, 9(3), 113; doi:10.3390/universe9030113
	
	\bibitem[Oates(2023)]{Oates2023} Oates, S.,
	\href{https://doi.org/10.3390/universe9030113}{2023, Universe, 9(3), 113}
	
	\bibitem[Paciesas et al.(2012)]{Paciesas2012} Paciesas W.~S., Meegan C.~A., von Kienlin A., et al., \href{https://doi.org/10.1088/0067-0049/199/1/18}{2012, ApJS, 199, 18}
	
	\bibitem[Perley et al.(2016)]{Perley_2016}Perley D.~A., Schulze S., de Ugarte Postigo A., et al., \href{https://doi.org/10.3847/0004-637X/817/1/7}{2016, ApJ, 817, 7}
	
	\bibitem[Phillips et al.(1993)]{Phillips_1993} Phillips M.~M. et al.
	\href{https://ui.adsabs.harvard.edu/abs/1993ApJ...413L.105P}{1993, ApJ, 413, L105}
	
	
	\bibitem[Planck Collaboration(2020)]{Planck_2020} Planck Collaboration, Aghanim N., Akrami Y., et al., \href{https://doi.org/10.1051/0004-6361/201833880}{2020, A\&A, 641, A1}
	
	
	\bibitem[Poolakkil et al.(2021)]{Poolakkil2021} Poolakkil S., Preece R., Fletcher C., et al., \href{https://doi.org/10.3847/1538-4357/abf24d}{2021, ApJ, 913, 60}
	
	
	\bibitem[Reichart et al.(2001)]{reichart2001dust} Reichart D.~E. \href{https://ui.adsabs.harvard.edu/abs/1999astro.ph.12368R}{arXiv: astro-ph/9912368}
	
	\bibitem[Riess et al.(2022)]{Riess2022} Riess, A.G., et al,
	\href{https://doi.org/10.3847/1538-4357/ac8f24}{2022, ApJ, 938, 36}
	
	
	\bibitem[Salvaterra et al.(2009)]{Salvaterra2009} Salvaterra R., Della Valle M., Campana S., et al.,  \href{https://doi.org/10.1038/nature08445}{2009, Nature, 461, 1258}
	
	\bibitem[Schaefer(2003)]{Schaefer2003} Schaefer, B. E.
	\href{https://doi.org/10.1086/368104}{2003, ApJ, 583, L67}
	\bibitem[Schaefer(2007)]{Schaefer2007} Schaefer, B. E.
	\href{https://doi.org/10.1086/511742}{2007, ApJ, 660, 16}
	
	\bibitem[Schroeder  et al. (2022)]{Schroeder2022}  Schroeder, G.,  Laskar, T.,  Fong, W.  et al.
	\href{https://doi.org/10.3847/1538-4357/}{2022, ApJ, 940, 53}
	
	\bibitem[Scolnic et al.(2018)]{Scolnic2018} Scolnic, D. M., Jones, D. O., Rest, A., et al., \href{https://doi.org/10.3847/1538-4357/aab9bb}{2018, ApJ, 859, 101}
	\bibitem[Scolnic et al.(2022)]{Scolnic2022} Scolnic D., Brout D., Carr A., et al.,
	\href{https://doi.org/10.3847/1538-4357/ac8b7a}{2022, ApJ, 938, 113}
	
	
	\bibitem[Seikel et al.(2012)]{seikel_2012} Seikel M., Clarkson C., \& Smith M. \href{https://doi.org/10.1088/1475-7516/2012/06/036}{2012, JCAP, 06, 036}
	
	\bibitem[Tanvir et al.(2009)]{Tanvir2009} Tanvir N.~R., Fox D.~B., Levan A.~J., et al., \href{https://doi.org/10.1038/nature08459}{2009, Nature, 461, 1254}
	
	\bibitem[Tang et al.(2019)]{Tang_2019} Tang C. H., Huang Y. F., Geng J. J., \& Zhang Z. B. \href{https://doi.org/10.3847/1538-4365/ab4711}{2019, ApJS, 245, 1}
	
	\bibitem[Tang et al.(2021)]{Tang2021} Tang L., Li X., Lin, H.-N., \& Liu L.
	\href{https://doi.org/10.3847/1538-4357/abcd92} {2021, ApJ, 907, 121}
	
	\bibitem[Tsvetkova et al.(2021)]{Tsvetkova2021}  Tsvetkova A.,  Frederiks D.,  Svinkin, D.  et al.,
	\href{https://doi.org/10.3847/1538-4357/abd569} {2021, ApJ, 908, 83}
	
	\bibitem[Von~Kienlin et al.(2014)]{von2014second} von Kienlin A., Meegan C.~A., Paciesas W.~S., et al., \href{https://doi.org/10.1088/0067-0049/211/1/13}{2014, ApJS, 211, 13}
	
	\bibitem[Von~Kienlin et al.(2020)]{von2020fourth} von Kienlin A., Meegan C.~A., Paciesas W.~S., et al., \href{https://doi.org/10.3847/1538-4357/ab7a18}{2020, ApJ, 893, 46}
	
	\bibitem[Wang \& Dai(2006)]{Wang2006} Wang, F., \& Dai, Z. G.
	\href{https://doi.org/10.1111/j.1365-2966.2006.10108.x}{2006, MNRAS, 368, 371}
	\bibitem[Wang et al.(2016)]{Wang2016} Wang, J. S., Wang, F. Y., Cheng, K. S., \& Dai, Z. G. \href{https://doi.org/10.1051/0004-6361/201526485}{2016, AA, 585, A68}
	\bibitem[Wang et al.(2024)]{Wang2024} Wang, G.Z.,  Li, X. L. \& Liang, N.
	\href{} {arXiv:2404.14237}
	\bibitem[Wei et al.(2016)]{Wei_2016} Wei J., Cordier B., Antier S., et al.,
	\href{https://doi.org/10.48550/arXiv.1610.06892}{2016, arXiv:1610.06892}
	\bibitem[Wei \& Zhang(2009)]{Wei2009} Wei, H., \& Zhang, S. N.
	\href{https://doi.org/10.1140/epjc/s10052-009-1086-z}{2009, EPJC, 63, 139}
	\bibitem[Wei(2010)]{Wei2010} Wei, H.,
	\href{https://doi.org/10.1088/1475-7516/2010/08/020}{2010, JCAP, 08, 020}
	
	
	
	\bibitem[Xie et al.(2023)]{Xie2023} Xie, H., Nong, X., Wang, H., Zhang, B., Li, Z. \& Liang, N.
	\href{} {arXiv:2307.16467}
	\bibitem[Yonetoku et al.(2004)]{Yonetoku_2004} Yonetoku D., Murakami T., Nakamura T., et al., \href{https://doi.org/10.1086/421285}{2004, ApJ, 609, 935}
	\bibitem[Zhang et al.(2023)]{Zhang2023} Zhang, B., Xie, X., Nong, X., Wang, H.,  Li, Z. \& Liang, N.
	\href{} {arXiv:2312.09440}
	\bibitem[Zhang, Huang, \& Zou.(2023)]{ZHZ2023} Zhang X.-L., Huang Y.-F., Zou Z.-C., \href{https://doi.org/10.1088/1674-4527/acf18d}{2023, RAA, 23, 125003}
	
	\makeatother
\end{thebibliography}
\end{document}